\documentclass[entropy,article,accept,moreauthors,pdftex]{Definitions/mdpi}

\firstpage{1}
\pubvolume{22}
\issuenum{9}
\articlenumber{1001}
\pubyear{2020}
\copyrightyear{2020}
\history{Received: 16 August 2020; Accepted: 4 September 2020; Published: 8 September 2020}

\usepackage{todonotes,bm}
\usepackage[normalem]{ulem}
\usepackage{mathabx}
\usepackage{scalerel}
\usetikzlibrary{arrows.meta}
\definecolor{ashgrey}{rgb}{0.7, 0.75, 0.71}
\usepackage{cancel}

\usepackage[labelformat=simple]{subcaption}

\DeclareCaptionLabelFormat{subcaptionlabel}{\normalfont(\textbf{#2}\normalfont) }
\Title{Flow Statistics in the Transitional Regime of Plane Channel Flow}

\Author{Pavan V. Kashyap $^{1,}$*, Yohann Duguet $^{1}$\href{https://orcid.org/0000-0001-6258-0475}{\orcidicon} and Olivier Dauchot $^{2}$\href{https://orcid.org/0000-0002-7039-5787}{\orcidicon}}

\AuthorNames{Pavan V. KASHYAP, Yohann DUGUET and Olivier DAUCHOT}

\address{%
$^{1}$ \quad LIMSI-CNRS, UPR 3251, Universit\'e Paris-Saclay, 91405 Orsay, France; yohann.duguet@limsi.fr\\
$^{2}$ \quad Gulliver, ESPCI-CNRS, 10 Rue Vauquelin, 75005 Paris, France; olivier.dauchot@espci.fr}

\corres{Correspondence: pavan.kashyap@limsi.fr}

\abstract{The transitional regime of plane channel flow is investigated {above} the transitional point below which turbulence is not sustained, using direct numerical simulation in large domains. Statistics of laminar-turbulent spatio-temporal intermittency are reported. The geometry of the pattern is first characterized, including statistics for the angles of the laminar-turbulent stripes observed in this regime, with a comparison to experiments. High-order statistics of the local and instantaneous bulk velocity, wall shear stress and turbulent kinetic energy  are then provided. The~distributions of the two former quantities have non-trivial shapes, characterized by a large kurtosis and/or skewness. Interestingly, we observe a strong linear correlation between their kurtosis and their skewness squared, which is usually reported at much higher Reynolds number in the fully turbulent regime.}


\begin{document}

\section{Introduction}
Laminar and turbulent flows are two different regimes {encountered sometimes at the same parameters for a given geometry}.~In many flows they are in competition from the point of view of the state space. Shear flows next to solid walls however show this surprisingly robust property that both laminar and turbulent regions coexist spatially on very long time scales, when the laminar state is locally stable. This phenomenon, called 'laminar-turbulent intermittency' is well known in circular pipe flow since the days of O. Reynolds \cite{Reynolds1883iii} and has lead recently to a burst of interest, a~review of which is provided in Reference \cite{Tuckerman2020arfm}. Such laminar-turbulent flows have been identified and partly characterized in Taylor-Couette flow \cite{Coles1965transition,Prigent2002large} and in plane Couette flow \cite{Prigent2002large,Barkley2005computational,Duguet2010formation}.~They also have been identified in other set-ups involving curvature \cite{Cros2002spatiotemporal,Ishida2016transitional,Kunii2019} or stabilizing effects \cite{brethouwer2012turbulent}. The transitional regimes of plane Poiseuille flow, the flow between two fixed parallel plates driven by a fixed pressure gradient, have not received as much attention although this flow is the archetype of wall-bounded turbulent flows. Although this flow is frequently cited as an example of flow developing a linear instability (under the form of Tollmien–Schlichtling waves) \cite{Orszag1971accurate}, coherent structures typical of laminar-turbulent coexistence have been  frequently reported in channel flow well below the linear instability threshold and  a series of experimental and cutting-edge numerical studies in the 1980s and 1990s have focused on the development of spots \cite{Carlson1982flow,Alavyoon1986turbulent, Henningson1987wave,li1989wave,HenningsonKim1991spots}. Sustained intermittent regimes have not been identified as such before the mid-2000s, when Tsukahara \cite{Tsukahara2005dns} reported large-scale coherent structures from numerics in larger numerical domains. Like their counterpart in Couette flows, these structures display obliqueness with respect to the mean flow direction and a complicated long-time dynamics. The dynamics at onset in particular have remained mysterious \cite{Tao2018extended} and, although this is currently debated, could follow a~scenario different from the directed percolation one proposed for Couette flow. \cite{Lemoult2016directed,Chantry2017universal,Kunii2019}. In recent years, the so-called transitional regime of plane channel flow has attracted renewed attention after new experimental studies. Although the works in Refs \cite{Lemoult2012experimental,Lemoult2013turbulent,Lemoult2014turbulent} focused on the minimal transition amplitude for spot development, other studies \cite{Hashimoto2009experimental,seki2012experimental,Sano2016universal,paranjape2019thesis,whalley2019experimental,agrawal2020investigating} focused on the sustained intermittent regimes and their statistical quantification.

Experimentally the finite length of the channel sets a limitation to most statistical approaches. Numerical simulation in large domains combined with periodic boundary conditions is a well-established way to overcome such limitations. Surprisingly, despite a large number of numerical studies of transitional channel flow, investigation of spatio-temporal intermittency in large enough domains has not been possible before the availability of massive computational resources. Owing to recent numerical studies \cite{Xiong2015turbulent, Tuckerman2014turbulent,gome2020statistical}, there is currently a good consensus about a few facts concerning the transitional regime:  laminar-turbulent bands with competing orientations emerge progressively as the Reynolds number is reduced below $Re_{\tau} \approx 100$, and their mean wavelength increases as the Reynolds number is decreased. At even lower flow rate the bands turn into isolated spots with ballistic dynamics rather than forming a seemingly robust stripe pattern \cite{Kanazawa2018thesis,shimizu2019bifurcations,xiao2020growth}. The~global centerline Reynolds number for the disappearance of the stripes is close to 660 \cite{Tao2018extended,paranjape2019thesis}. However, many questions remain open. The~most sensible theoretical issues revolve around the (still open) question of the universality class of the transition process (see Reference \cite{Tao2018extended}),  the role of the large-scale flows \cite{seki2012experimental,Duguet2013oblique,Lemoult2014turbulent,couliou2015large} in the sustainment of the stripes, or the mutual way different stripes interact together.

There is also a lack of quantitative data about the patterning regime itself. The present special issue is an opportunity to document the geometric characteristics of the stripe patterns in unconstrained settings. Moreover, there is an ongoing philosophical question about whether traces of spatio-temporal intermittency can be found in the fully turbulent regimes commonly reported at higher Reynolds numbers. In the present paper, using numerical simulation in large domains, we focus on three specific points hitherto undocumented: the angular distribution of turbulent stripes, the statistics of the laminar gaps between them, and high-order statistics of the local and instantaneous bulk velocity, wall shear stress and turbulent kinetic energy. The outline of the paper is as follows: Section~\ref{sm} introduces the numerical methodology with the relevant definitions. The geometrical statistics of the stripe angles are presented in Section~\ref{sang}. The statistics of a few global quantities are presented in Sections~\ref{scf}--\ref{stm}. A~discussion of the results is made in Section~\ref{sdis} with the conclusions and outlooks in Section~\ref{scon}.

\section{Materials and Methods} \label{sm}

The present section is devoted to the methodology used for the numerical simulation of pressure-driven plane channel flow.~The flow is governed by the incompressible Navier Stokes equations.~Channel flow is described here using the Cartesian coordinates $x$,$y$,$z$, respectively the streamwise, wall-normal and spanwise coordinates. The velocity field ${\bm u}(x,y,z,t)$ is decomposed into the steady laminar base flow solution $\mathbf{U}(y)=(U_x,0,0)$ and a perturbation field ${\bm u^{\prime}(x,y,z,t)}$. Similarly, the pressure field is decomposed as $p(x,y,z,t)=xG +p^{\prime}(x,y,z,t)$. The equation governing the steady base flow for an incompressible fluid with constant density $\rho$ and kinematic viscosity $\nu$ is given by
\begin{equation}
\nu \frac{\partial^2 U_x}{\partial y^2} = \frac{1}{\rho} G
\label{Ubase}
\end{equation}
with $G$ a constant. Together with the no-slip condition at the walls Equation (\ref{Ubase}) yields the analytic Poiseuille solution $U_x \propto 1-(y/h)^2$. The equation governing the perturbation field involves the base flow and reads

\begin{equation}
\frac{\partial \mathbf{u}^\prime}{\partial t} + \mathbf{u^\prime} \cdot \nabla \mathbf{u^\prime} + \mathbf{U} \cdot \nabla \mathbf{u^\prime} + \mathbf{u^\prime} \cdot \nabla \mathbf{U} = - \frac{1}{\rho} \nabla p^\prime + \nu \nabla^2 \mathbf{u}^\prime \label{NS}
\end{equation}

\begin{figure}[H]
\centering

\begin{tikzpicture}[scale=0.35]

\coordinate (A1) at (-12.5,-1,-12.5);
\coordinate (A2) at (-12.5,-1,0);
\coordinate (A3) at (12.5,-1,0);
\coordinate (A4) at (12.5,-1,-12.5);
\coordinate (B1) at (-12.5,1,-12.5);
\coordinate (B2) at (-12.5,1,0);
\coordinate (B3) at (12.5,1,0);
\coordinate (B4) at (12.5,1,-12.5);

\draw[fill=ashgrey,opacity=1] (A1)--(A2)--(A3)--(A4)--cycle;
\draw[fill=ashgrey,opacity=1] (B1)--(B2)--(B3)--(B4)--cycle;

\draw[domain=-1:1,variable=\y,samples=50,line width=2,red] plot(2-2*\y*\y,\y);

\draw[-{Latex[scale=0.75]},yellow] (0,-0.6,0) -- (1.28,-0.6,0);
\draw[-{Latex[scale=0.75]},yellow] (0,-0.2,0) -- (1.92,-0.2,0);
\draw[-{Latex[scale=0.75]},yellow] (0,0.6,0) -- (1.28,0.6,0);
\draw[-{Latex[scale=0.75]},yellow] (0,0.2,0) -- (1.92,0.2,0);

\draw[black,thin,-{Latex[scale=1.0]}] (0,0,0)--(0,5,0)node[anchor=north west]{\fontsize{10pt}{10pt} $y$};
\draw[black,thin,-{Latex[scale=1.0]}] (0,0,0)--(0,-3,0);
\draw[black,thin,-{Latex[scale=1.0]}] (0,0,0)--(15,0,0)node[anchor=north west]{\fontsize{10pt}{10pt} $x$};
\draw[black,thin,-{Latex[scale=1.0]}] (0,0,0)--(0,0,5)node[anchor=north east]{\fontsize{10pt}{10pt} $z$};

\draw[black,-,thin] (B2) -- (-12.5,3,0);
\draw[black,-,thin] (B3) -- (12.5,3,0);
\draw[{Latex[scale=1.0]}-{Latex[scale=1.0]},black,thin] (-12.5,2.5,0) -- (12.5,2.5,0);
\draw[black,thin,-] (B2) -- (-12.5,3,0);
\draw[black,thin,-] (B1) -- (-12.5,3,-12.5);
\draw[{Latex[scale=1.0]}-{Latex[scale=1.0]},black,thin] (-12.5,2.5,0) -- (-12.5,2.5,-12.5) ;
\draw[black,thin,-] (-12.5,-1,0) -- (-15,-1,0);
\draw[black,thin,-] (-12.5,1,0) -- (-15,1,0);
\draw[{Latex[scale=1.0]}-{Latex[scale=1.0]},black,thin] (-14,-1,0) -- (-14,1,0) ;

\node (n1) at (-3,3.2,0){\fontsize{10pt}{10pt} $L_x$};
\node (n1) at (-13.5,3,-5){\fontsize{10pt}{10pt} $L_z$};
\node (n1) at (-15,0,0){\fontsize{10pt}{10pt} $2h$};
\end{tikzpicture}
\caption{Schematic of the numerical domain with the laminar base flow profile (red).}
\label{smd}
\end{figure}
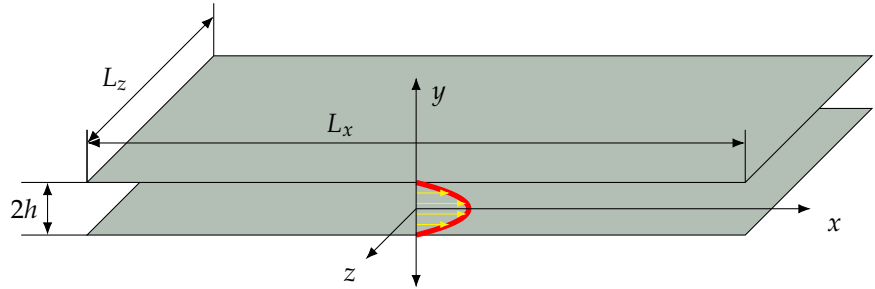

The channel geometry is formally infinitely extended, yet in the numerical representation it is given by its extent $L_x \times 2h \times L_z$ as in Figure \ref{smd}, with stationary walls at $y=\pm h$ and periodic boundary conditions in $x$ and $z$.

The flow is driven by the imposed pressure gradient $G$ assumed negative. The spanwise pressure gradient is explicitly constrained to be null. The centerline velocity $u_{cl}$ of the laminar base profile  with the same pressure gradient is chosen as the velocity scale ($U$) and the half gap $h$ of the channel is chosen as the lengthscale used for non-dimensionalization. Time is hence expressed in units of $h/U$.
In these units the laminar velocity profile is given by $U^{*}_x(y_{*}) = 1-y_{*}^2$. From Chapter 3 onwards only dimensionless quantities will be used and the $*$ notation will be dropped {from} there on. Primed quantities denote perturbations to the base flow while non-primed quantities involve the full velocity field, including the laminar base flow.

In the following we shall consider, both locally and temporally fluctuating quantities, as well as their time and space averages. We denote by $\left< \bullet \right>$ the space ($x,z$) average and $\widebar{\bullet}$---the time average. Space-time averages are indicated by $\widebar{ \left< \cdot \right>}$. More explicitly the space-average operator is defined as the discrete average over the grid points, and the time average is the discrete average sum over the total number of snapshots in the steady regime.

Different velocity scales characterize the flow. One such scale is the centerline velocity $u_{cl}$ of the corresponding laminar flow with the same value of $G$. Another one is the total streamwise flow through the channel, $U_b = \widebar{ \left< u_b\right> }$, where
\begin{equation}
u_b(x,z,t) = \frac{2}{h} \int_{-h}^{h} u_x dy
\end{equation}
is the so-called local bulk flow. Finally, the friction velocity is defined as $U_{\tau} = ( \widebar{ \left< \tau \right> } / \rho)^{\frac{1}{2}}$, where $ \tau = \left( \tau_t + \tau_b \right)/2>0$, with $\tau_t$ and $\tau_b$  the net shear stress on the top and the bottom wall, respectively given by:
\begin{equation}
\tau_{t,b} (x,z,t) = \pm \mu \frac{\partial u_x}{\partial y}\bigg|_{t,b} \label{sh}
\end{equation}
where $\mu=\rho \nu$ is the dynamic viscosity of the fluid. The  three Reynolds numbers arising from these velocity scales are $Re_{cl} = u_{cl} h/\nu$, $Re_b = U_b h/\nu$ and $Re_{\tau} = U_{\tau} h / \nu$. For the laminar base flow, they are inter-related as $Re^2_{\tau} = 3 Re_b = 2 Re_{cl}$. Imposing a pressure gradient $G$ < 0 translates into a fixed average shear stress $\widebar{ \left< \tau \right>}$ on the walls which sets an imposed value of $Re_{\tau} = Re^{\scaleto{G}{4pt}}_{\tau}$ to stress that this is the control~parameter.

Direct numerical simulation (DNS) of Equation (\ref{NS}) is carried out using the open source, parallel solver called Channelflow \cite{Channelflow,channelflow2} written in C++. It is based on a Fourier–Chebychev discretization in space and a 3{rd} order semi-implicit backward difference scheme for timestepping. It makes use of the $2/3$ dealiasing rule for the nonlinear terms. An influence matrix method is used to ensure the no-slip boundary condition at the walls. The numerical resolution is specified in terms of the spatial grid points $(N_x,N_y,N_z)$ which translates into a maximum of $(N_x/2+1,N_z/2+1)$ Fourier wavenumbers and $N_y$ Chebychev modes. Please note that the definitions of $N_x$ and $N_z$ take into account the aliasing modes. The domain sizes used in this study, expressed in units of $h$, are $L_x=2L_z=250$ for $55<Re^{\scaleto{G}{4pt}}_{\tau} \leq 100$ and $L_x=2L_z=500$ for $39 \leq Re^{\scaleto{G}{4pt}}_{\tau} \leq 55$. The local numerical resolution used is $N_x/L_x=N_z/(2L_z)= 4.096$ and $N_y=65$, comparable to that used in Reference \cite{shimizu2019bifurcations}.
The simulation follows an ``adiabatic descent'': a first simulation is carried out at sufficiently high value of $Re^{\scaleto{G}{4pt}}_{\tau}$, known to display space-filling turbulence. After the stationary turbulent regime is reached, $Re^{\scaleto{G}{4pt}}_{\tau}$ is lowered and the simulation advanced further in time. This step-by-step reduction has been performed down to $Re^{\scaleto{G}{4pt}}_{\tau}=39$. The initial condition for the {first} simulation is a random distribution of localized seeds of the kind described in Reference~\cite{lundbladh1991direct}. The time required $T$ to reach a stationary regime gradually increases as $Re^{\scaleto{G}{4pt}}_{\tau}$ is decreased. As an order of magnitude, for $Re^{\scaleto{G}{4pt}}_{\tau}=100$, $T \approx 1500$, while for $Re^{\scaleto{G}{4pt}}_{\tau}=50$, $T \approx 3000$. Statistics are computed, after excluding such transients, from time series of lengths up to $2\times10^4$ time units.

\begin{figure}
\centering
\scalebox{.9}[0.9]{\begin{tikzpicture}
\node[inner sep=0pt] (f1) at (0,0)
{\hypertarget{fig2a}{\includegraphics[scale=0.37]{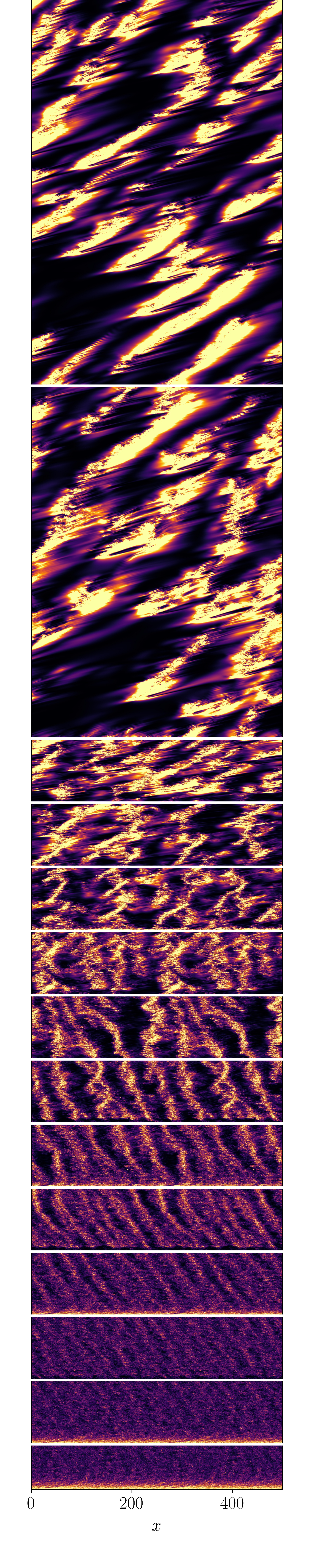}}};
\node[inner sep=0pt] at (0,-11.5) {\fontsize{10pt}{10pt} (\textbf{a})};
\node[inner sep=0pt] (fc) at (2.8,0)
{\includegraphics[scale=1]{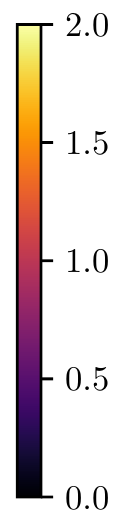}};
\node[inner sep=0pt] (fc1) at (2.8,3){\fontsize{10pt}{10pt} \textbf{$\times 10^{-3}$}};
\node[inner sep=0pt] at (-2.4,-10.6) {\fontsize{10pt}{10pt} $100$};
\node[inner sep=0pt] at (-2.4,-9.7) {\fontsize{10pt}{10pt} $95$};
\node[inner sep=0pt] at (-2.4,-8.8) {\fontsize{10pt}{10pt} $90$};
\node[inner sep=0pt] at (-2.4,-7.8) {\fontsize{10pt}{10pt} $85$};
\node[inner sep=0pt] at (-2.4,-6.8) {\fontsize{10pt}{10pt} $80$};
\node[inner sep=0pt] at (-2.4,-5.8) {\fontsize{10pt}{10pt} $75$};
\node[inner sep=0pt] at (-2.4,-4.8) {\fontsize{10pt}{10pt} $70$};
\node[inner sep=0pt] at (-2.4,-3.8) {\fontsize{10pt}{10pt} $65$};
\node[inner sep=0pt] at (-2.4,-2.8) {\fontsize{10pt}{10pt} $60$};
\node[inner sep=0pt] at (-2.4,-1.8) {\fontsize{10pt}{10pt} $55$};
\node[inner sep=0pt] at (-2.4,-0.9) {\fontsize{10pt}{10pt} $52$};
\node[inner sep=0pt] at (-2.4,0.2) {\fontsize{10pt}{10pt} $50$};
\node[inner sep=0pt] at (-2.4,3.5) {\fontsize{10pt}{10pt} $45$};
\node[inner sep=0pt] at (-2.4,7.7) {\fontsize{10pt}{10pt} $42$};

\node[inner sep=0pt] (f2) at (8.2,-9)
{\hypertarget{fig2b}{\includegraphics[width=0.55\textwidth]{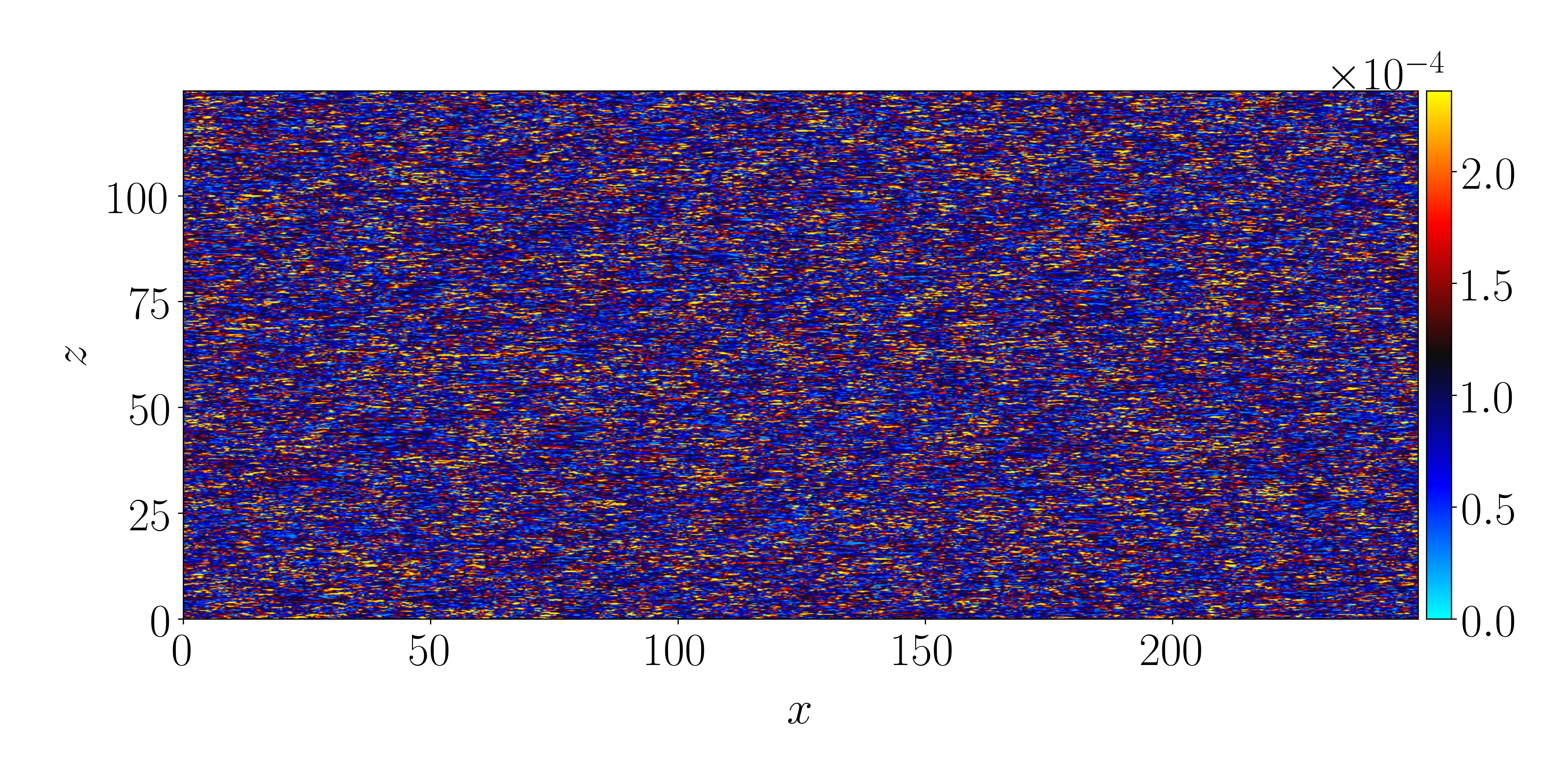}}};
\node[inner sep=0pt] at (8.2,-11.5) {\fontsize{10pt}{10pt} (\textbf{b})};
\node[inner sep=0pt] (f3) at (8.2,-2.5)
{\hypertarget{fig2c}{\includegraphics[width=0.55\textwidth]{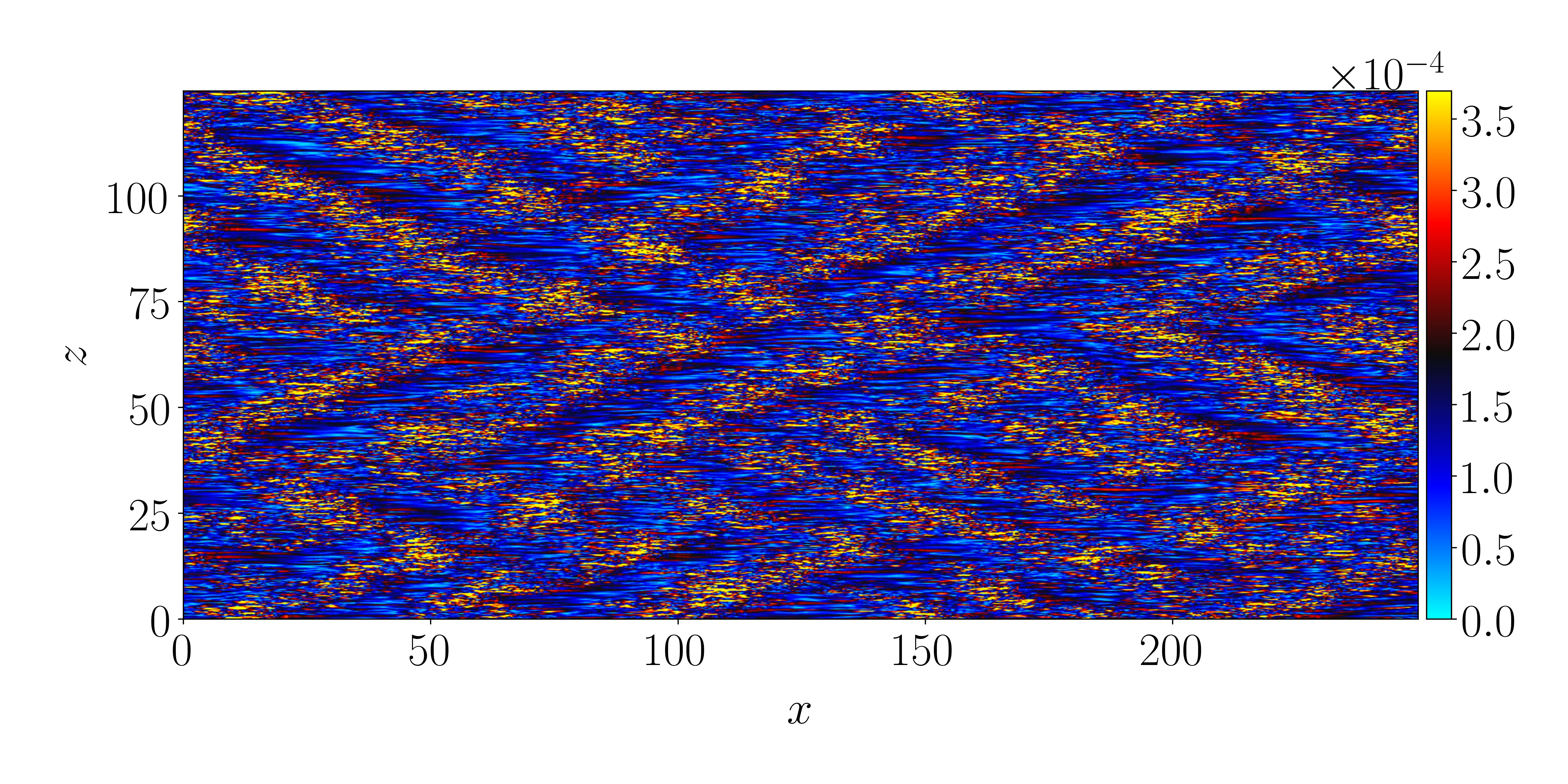}}};
\node[inner sep=0pt] at (8.2,-5) {\fontsize{10pt}{10pt} (\textbf{c})};
\node[inner sep=0pt] (f4) at (8.2,3.5)
{\hypertarget{fig2d}{\includegraphics[width=0.55\textwidth]{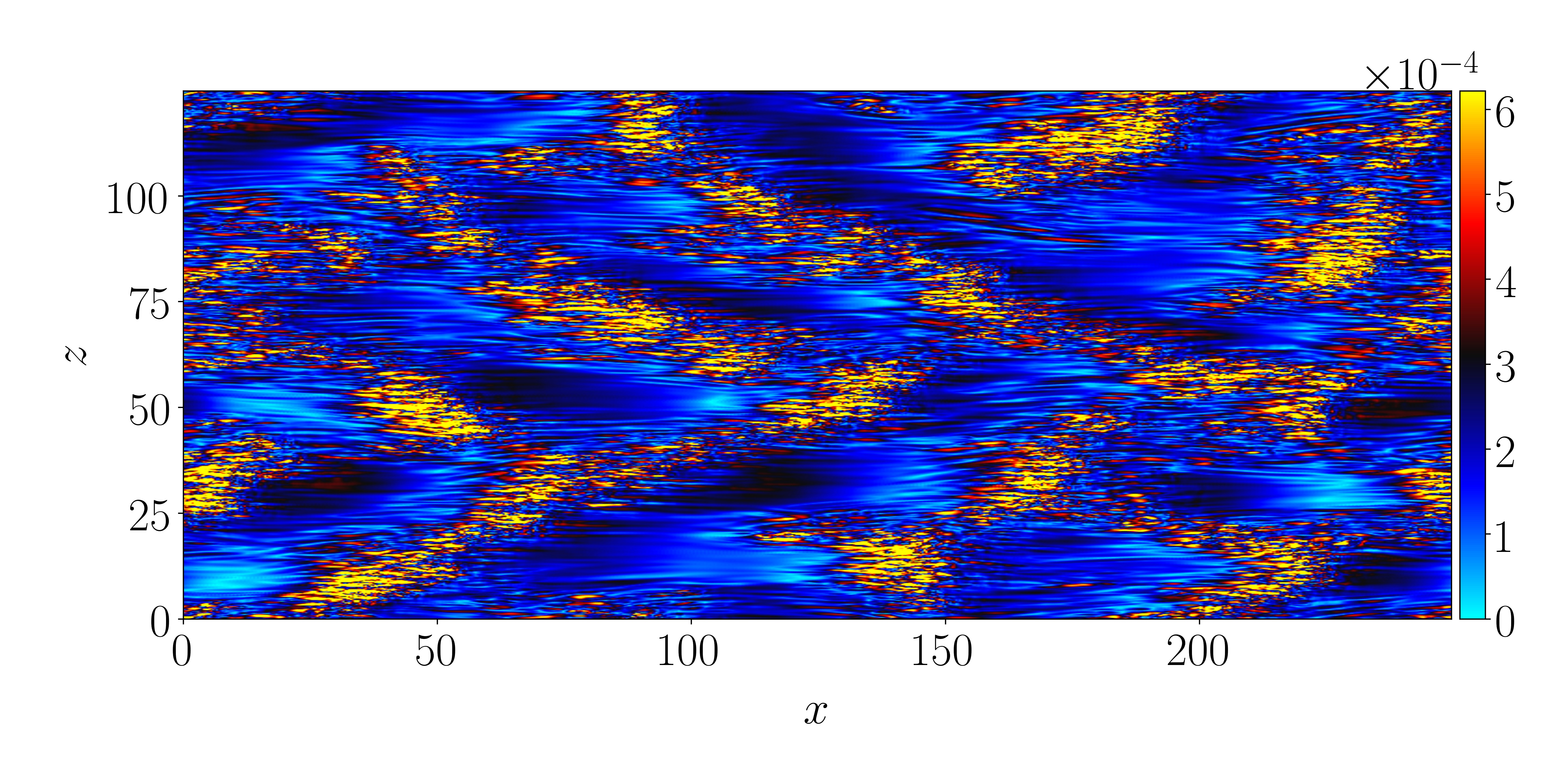}}};
\node[inner sep=0pt] at (8.2,1) {\fontsize{10pt}{10pt} (\textbf{d})};
\node[inner sep=0pt] (f5) at (8.2,9.5)
{\hypertarget{fig2e}{\includegraphics[width=0.55\textwidth]{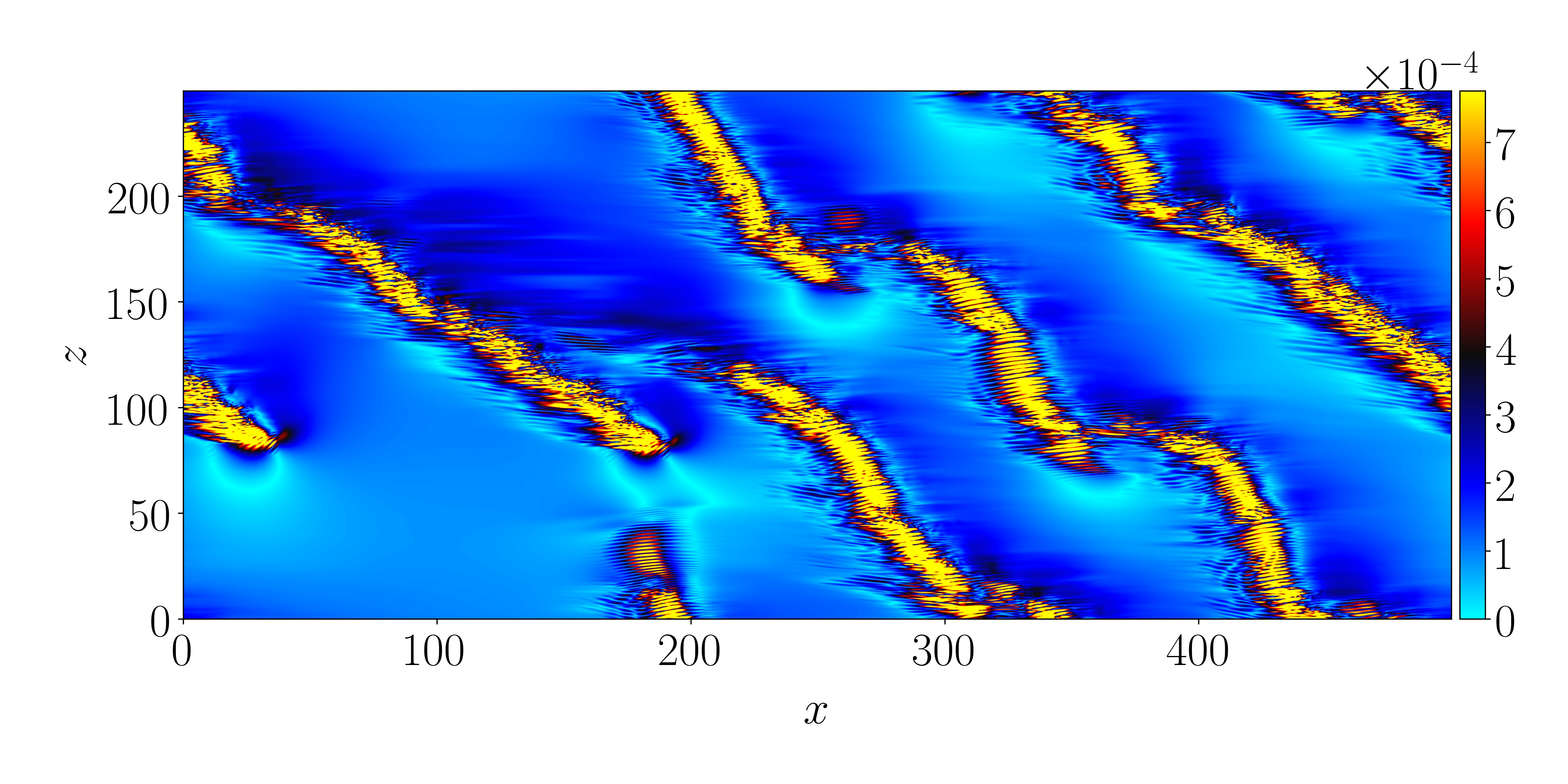}}};
\node[inner sep=0pt] at (8.2,7) {\fontsize{10pt}{10pt} (\textbf{e})};
\end{tikzpicture}}
\caption{({\bf a}) Space-time diagram of $E_{cf}(x-U_b(G) \, t,t)$ for $z=L_z/2$ during the adiabatic descent protocol, in a frame travelling in the $x$-direction at the mean bulk velocity $U_b(G)$. Vertical axis:  time with corresponding values of $Re^{\scaleto{G}{4pt}}_{\tau}$ values  indicated. ({\bf b--e}) isocontours of $\tau^\prime (x,z)$ for $Re^{\scaleto{G}{4pt}}_{\tau}=100,80,60,40$.}
\label{flowvis}
\end{figure}
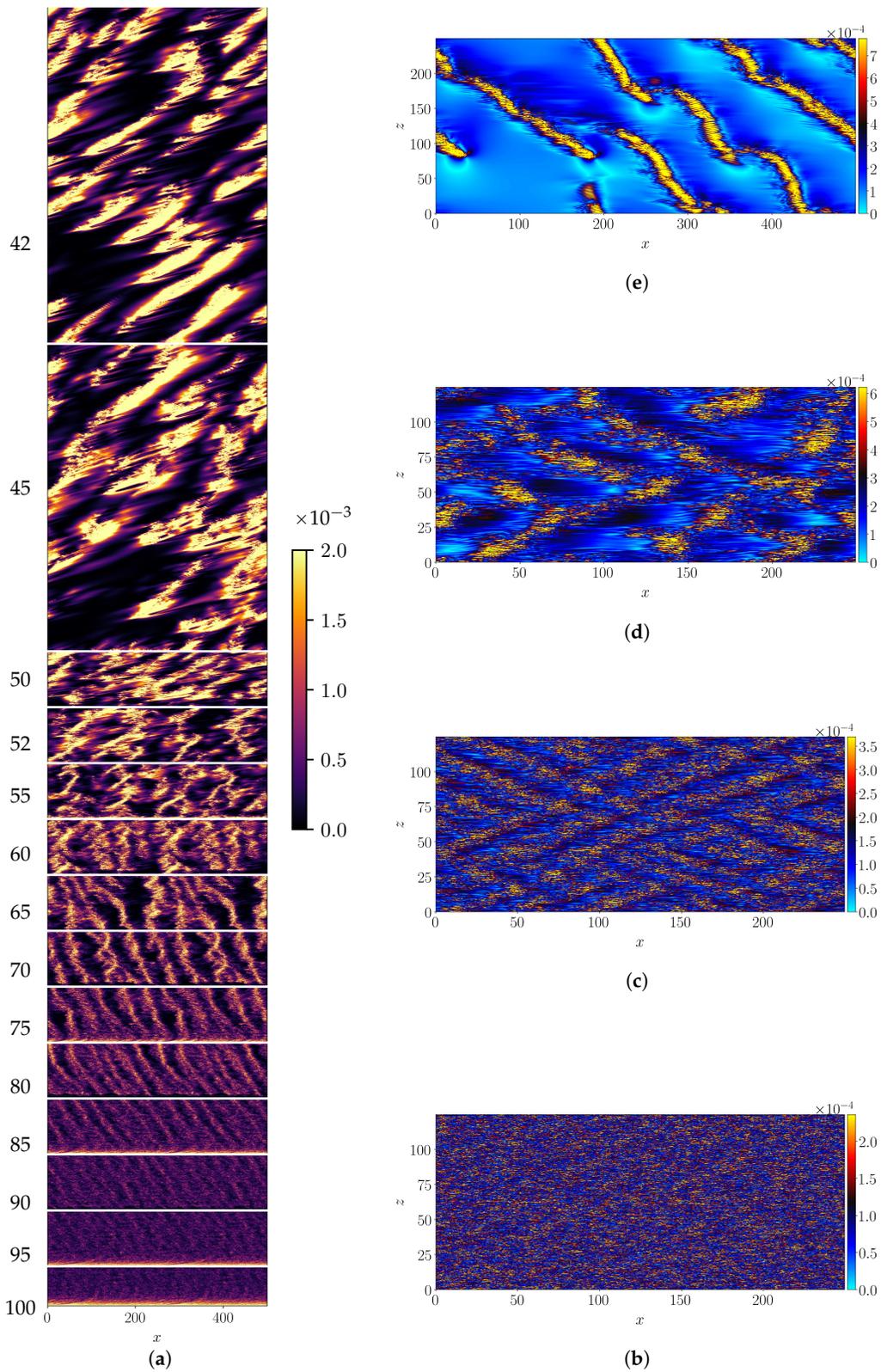

\section{Results} \label{sr}

The entire adiabatic descent is shown using a space-time diagram of the crossflow energy shown in Figure~\ref{flowvis}a 
\begin{equation}
E_{cf} = \frac{1}{2} \int  (u^2_y  + u^2_z) dy
\label{Ecf}
\end{equation}
evaluated at an arbitrary value of $z$ (here  $z=L_z/2$). The space variable is expressed in a frame moving in the streamwise direction with the mean bulk velocity $U_b (G)$ for that particular value of $Re^{\scaleto{G}{4pt}}_{\tau}$. Since $Re^{\scaleto{G}{4pt}}_{\tau}$ is lowered over the course of time, this allows one to capture the different flow regimes preceding full relaminarization. The intensity of turbulence, measured here by the value of $E_{cf}$, is seen to gradually increase as $Re^{\scaleto{G}{4pt}}_{\tau}$ is lowered. At high $Re^{\scaleto{G}{4pt}}_{\tau}$, the so-called featureless turbulence occupies the full domain, as shown in Figure~\ref{flowvis}{b} at $Re^{\scaleto{G}{4pt}}_{\tau}=100$ using isocontours of {$\tau^\prime(x,z)=\tau(x,z)-\tau_{lam}$}. As~$Re^{\scaleto{G}{4pt}}_{\tau}$ is lowered, turbulence self-organizes into the recognizable pattern regime \cite{Tsukahara2005dns} shown in Figure~\ref{flowvis}c for $Re^{\scaleto{G}{4pt}}_{\tau}=80$. As $Re^{\scaleto{G}{4pt}}_{\tau}$ is further reduced the turbulent zones become sparser (see Figure~\ref{flowvis}{d} for $Re^{\scaleto{G}{4pt}}_{\tau}=60$).  The spatially localized turbulent regions emerge as narrow stripes throughout the process of decreasing $Re^{\scaleto{G}{4pt}}_{\tau}$ while the gaps between them constantly increase in size. The emerging patterns never feature an array of strictly parallel stripes like in former computational approaches \cite{Tuckerman2014turbulent,Lemoult2016directed,paranjape2020oblique}, instead they feature competing orientations as in pCf \cite{Prigent2002large}, see Figure \ref{flowvis}{b}--{d}. In this regime the pattern travels with a streamwise convection velocity slightly slower than $U_b(G)$. Within the quasi-laminar gaps, $E_{cf}$~reaches very low values, at least an order of magnitude less than in the core of the turbulent stripes. The~lower $Re^{\scaleto{G}{4pt}}_{\tau}$, the lower these values. Below $Re^{\scaleto{G}{4pt}}_{\tau}=50$ the stripe pattern eventually breaks up to form independent turbulent bands of finite length, all parallel to each other \cite{shimizu2019bifurcations}, as shown in Figure~\ref{flowvis}{e} for $Re^{\scaleto{G}{4pt}}_{\tau}=40$. The new resulting pattern as a whole shows negligible spanwise advection, while it propagates in $x$ with a velocity close to $\widebar{\left<u_b\right>}$ \cite{fukudome2012large}. The independent turbulent bands show enhanced motility in both directions $x$ and $z$. This motion relative to the frame of reference causes the tilt of the stripes seen in Figure~\ref{flowvis}{a} for $Re^{\scaleto{G}{4pt}}_{\tau} > 50$ as well as the apparent increase of thickness.

\begin{figure}[H]
\centering
\includegraphics[width=0.5\textwidth]{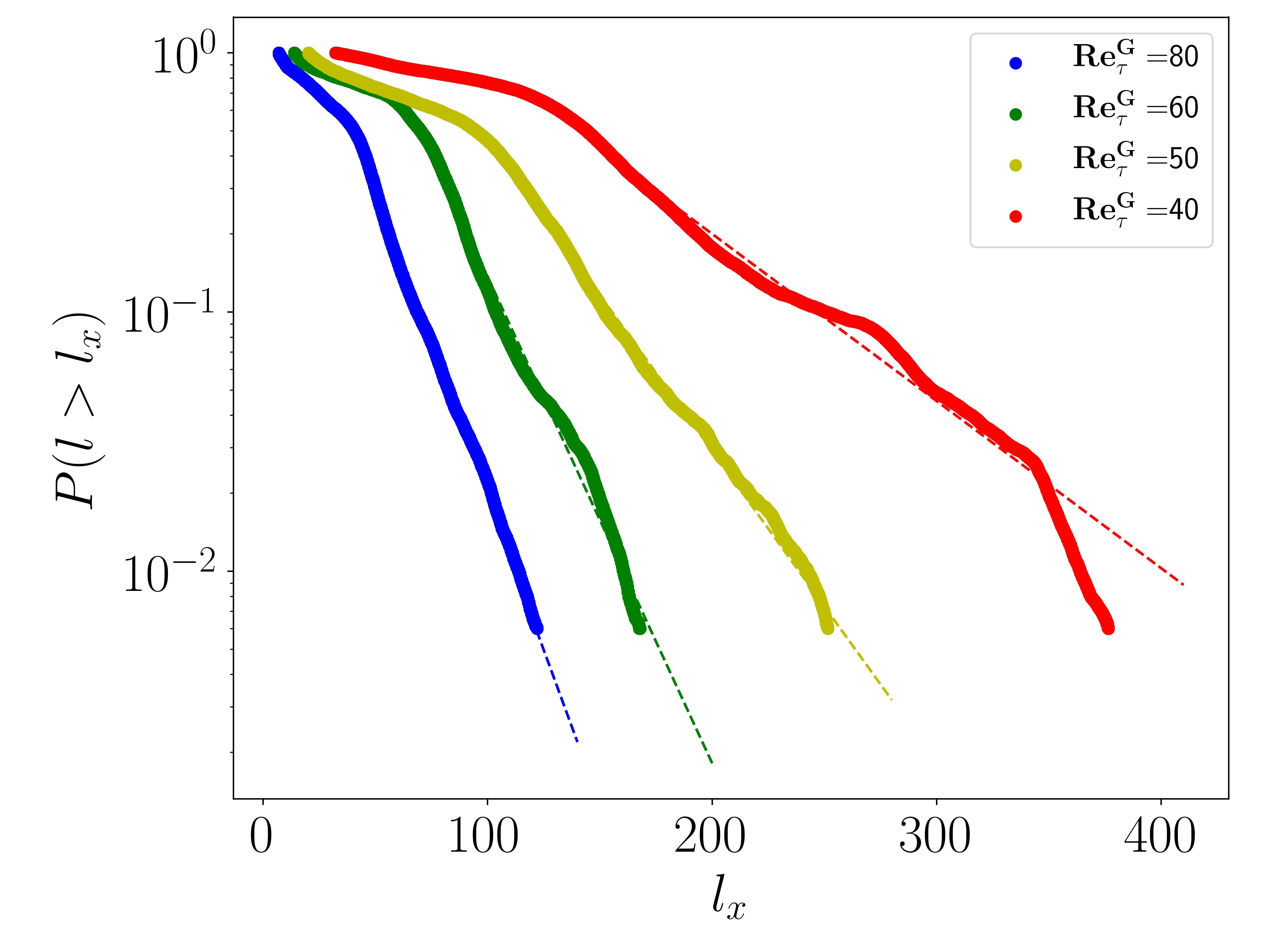}
\caption{ CDF of laminar gap size for $Re^{\scaleto{G}{4pt}}_{\tau}=80, 60, 50, 40$.}
\label{cdf}
\end{figure}

In pipe flow it was noted recently \cite{mukund2018critical} that the emergence of spatial localization does not imply the proximity to the transitional point (below which turbulence is not sustained) as long as the statistics about the size of the laminar gaps fail at displaying power-laws tails. The laminar gaps are estimated as the streamwise distance $l_x$ between local maxima of $\tau$ (values lower than $\left<\tau\right> + \sigma(\tau)$, with $\sigma$ the standard deviation, have been discarded).~The cumulative distribution (CDF) of the laminar gap size is shown in Figure \ref{cdf} in lin-log coordinates. For all values of $Re_{\tau}$ shown, it shows an exponential tails and no algebraic part. Exponential distributions are a hallmark of spatio-temporal intermittency, unlike critical phenomena which are characterized by algebraic/power law related to the scale invariance property. The entire regime of channel flow for $39 \leq Re^{\scaleto{G}{4pt}}_{\tau} \leq 100 $ can be described as being spatiotemporally intermittent, and is hence far {above} any critical point. Please note that the critical point of pPf is estimated to approximately $Re_{cl}=660$ \cite{Tao2018extended} i.e., $Re^{\scaleto{G}{4pt}}_{\tau} \approx 36$ and falls outside the range of parameters investigated here.


\subsection{Angular Statistics of Turbulent Bands} \label{sang}

\begin{figure}[H]
\centering
\begin{subfigure}[b]{0.55\textwidth}
\centering
\includegraphics[width=\textwidth]{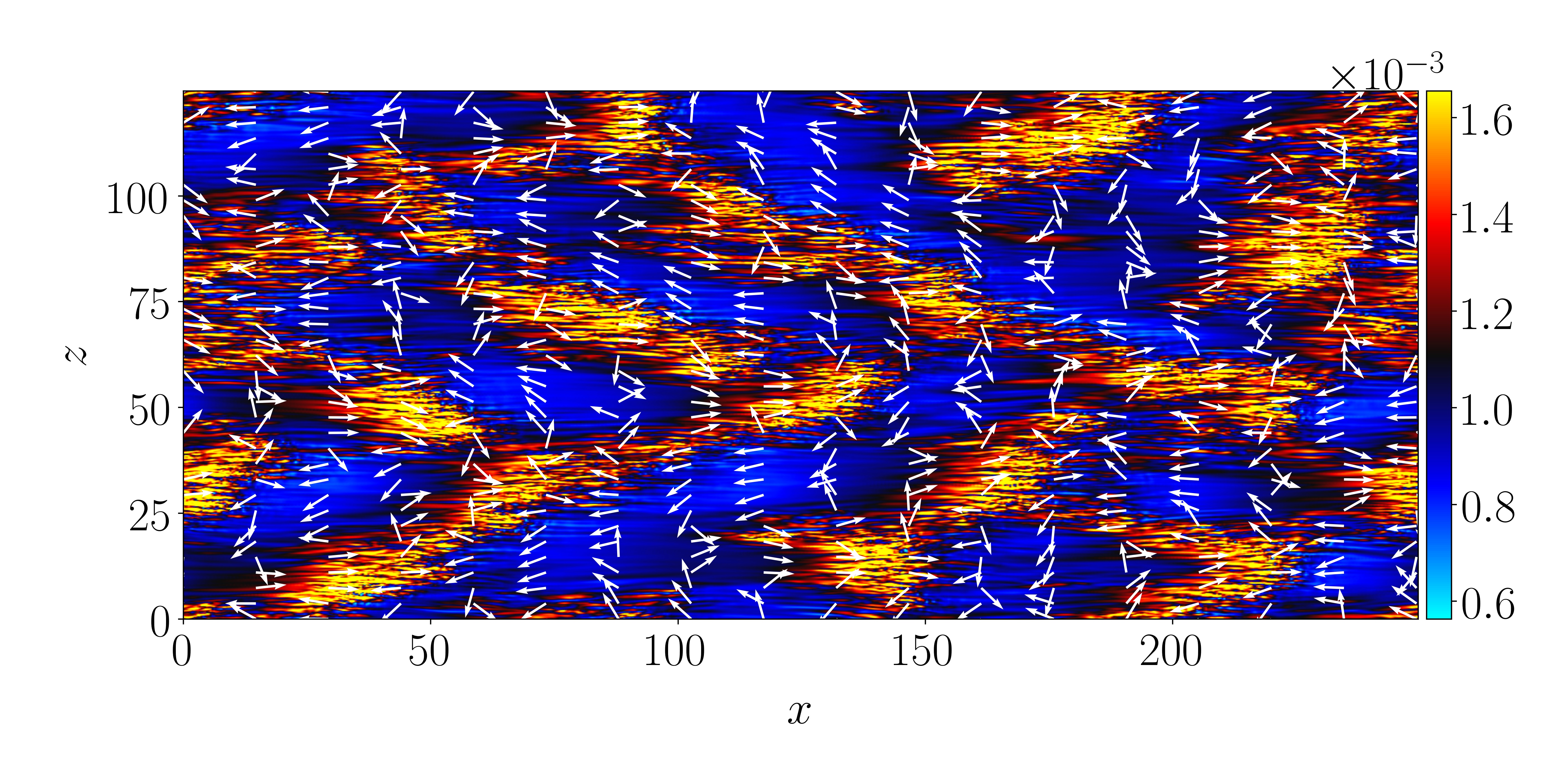}
\caption{}
\label{art60}
\end{subfigure}
\hfill
\begin{subfigure}[b]{0.35\textwidth}
\centering
\includegraphics[width=\textwidth,trim = 0.1cm 2cm 0.1cm 2cm,clip]{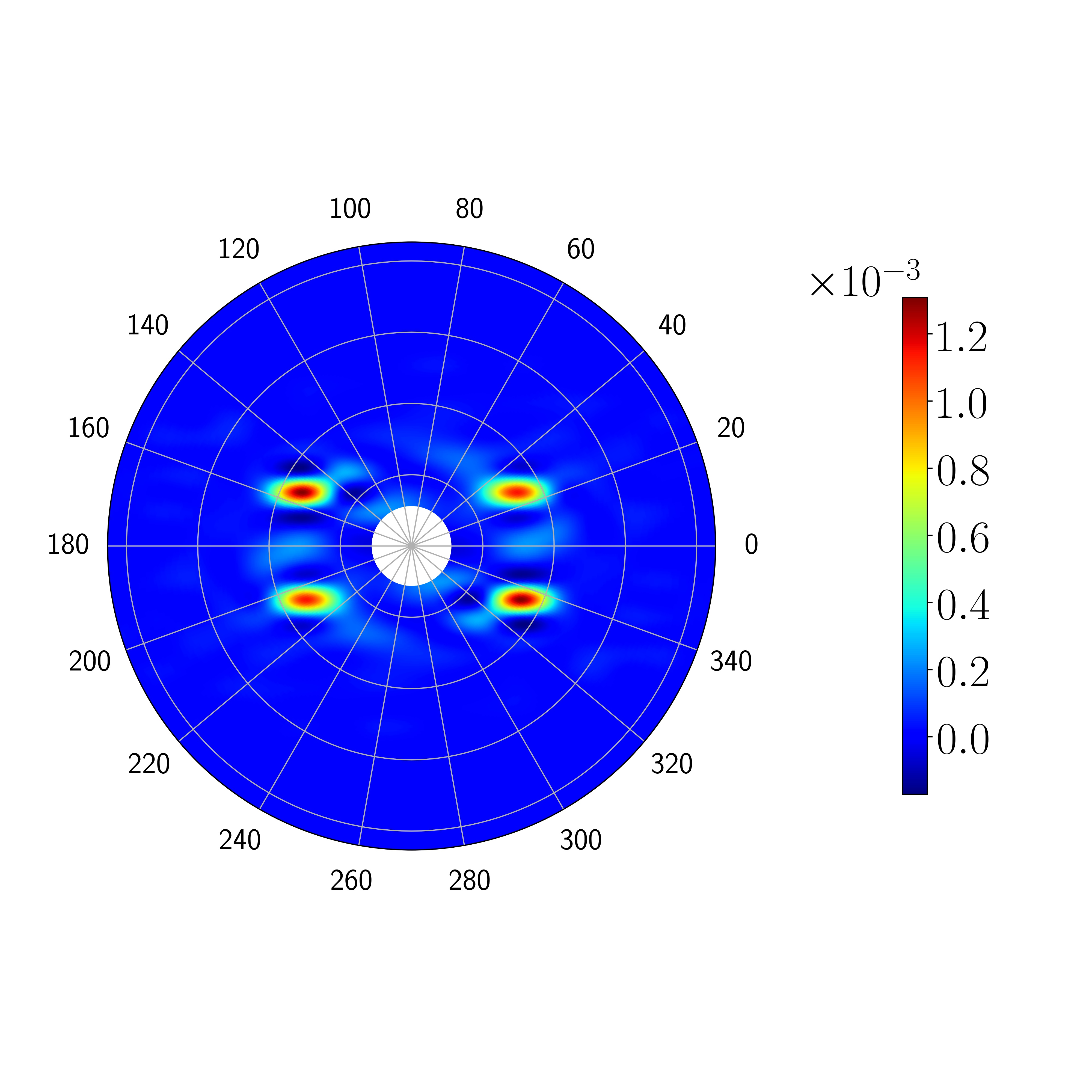}
\caption{}
\label{prt60}
\end{subfigure}
\hfill
\vspace{2mm}
\begin{subfigure}[b]{0.55\textwidth}
\centering
\includegraphics[width=\textwidth]{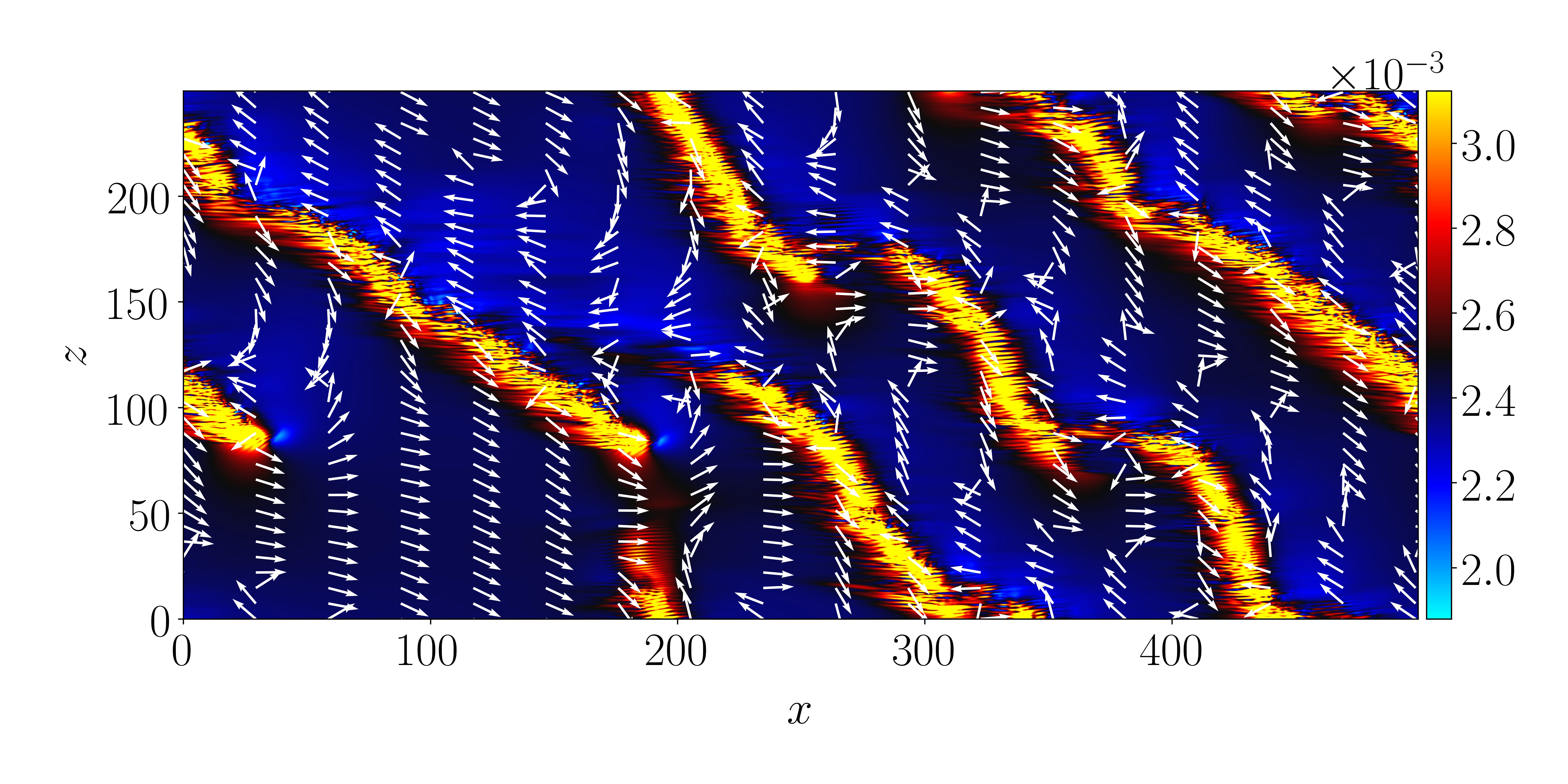}
\caption{}
\label{art40}
\end{subfigure}
\hfill
\begin{subfigure}[b]{0.35\textwidth}
\centering
\includegraphics[width=\textwidth,trim = 0.1cm 2cm 0.1cm 2cm,clip]{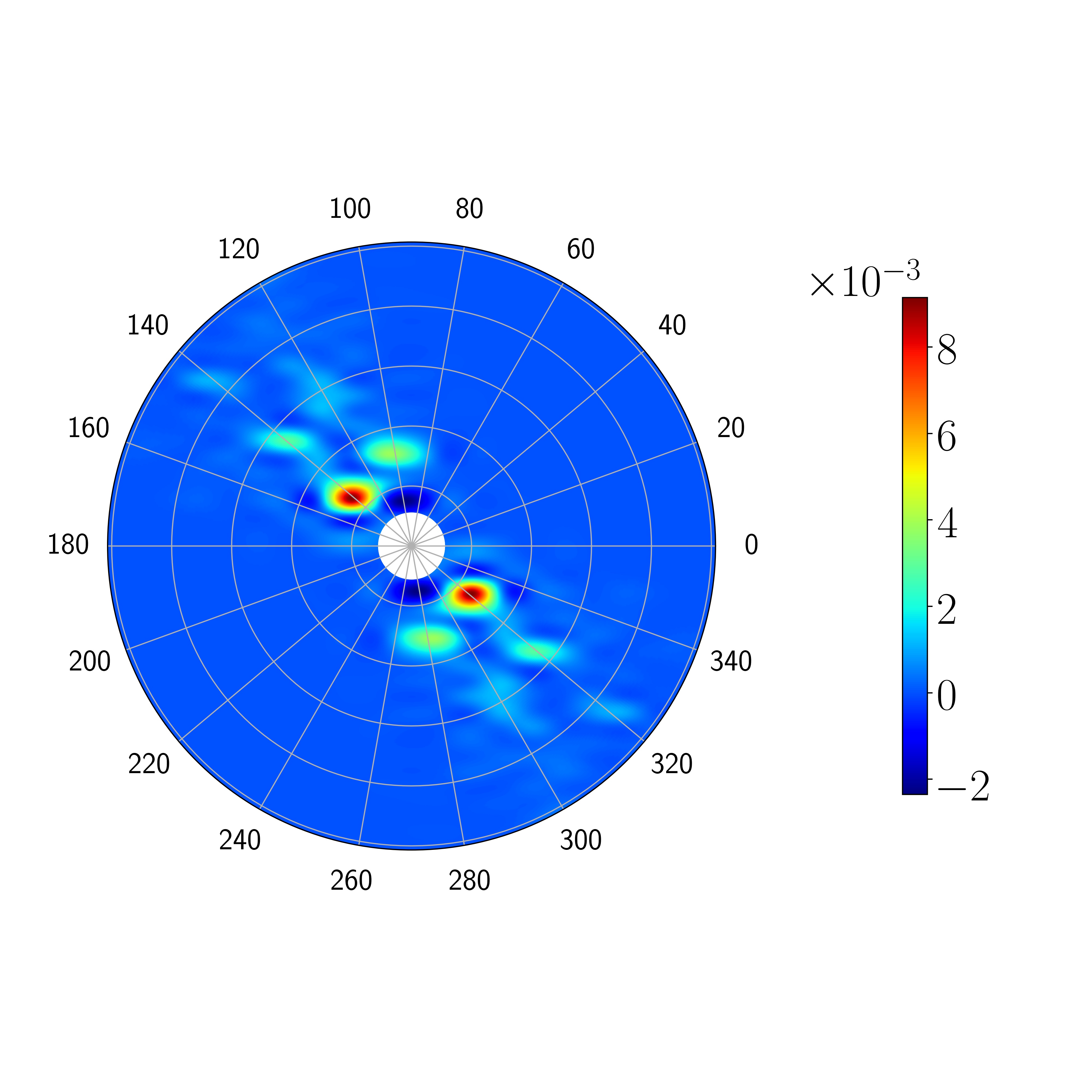}
\caption{}
\label{prt40}
\end{subfigure}
\hfill\hfill
\caption{({\bf a},{\bf c}) Isocontours of $\tau^\prime$ with the local velocity indicated by the normalized velocity vectors, at $Re^{\scaleto{G}{4pt}}_{\tau}=60,40$, respectively;  ({\bf b},{\bf d}) Instantaneous Fourier spectrum in polar coordinates for (\textbf{a},\textbf{b}),~respectively.}
\label{avis}
\end{figure}

The self-organization of turbulence into long band--like structures, oriented with an angle with respect to the streamwise direction, is depicted in \textls[-15]{Figure \ref{flowvis}. The (signed) angle is computed using two different methodologies. As in Duguet {et al.} \cite{Duguet2013oblique} in the case of pCf, the local $y$--integrated velocity field is found to be parallel to the bands. The same holds for pPf, as is visible in Figure \ref{avis}a,c for $Re^{\scaleto{G}{4pt}}_{\tau} = 60$ and $40$, respectively. Please note that unlike Couette flow, pPf features advection with a non-zero mean bulk velocity. Hence the local velocity field is here computed by removing this mean advection velocity. A first estimation of the local and instantaneous band angle is therefore computed following Equation (\ref{tl}):}
\begin{equation}
\theta_L (x,z,t) = tan^{-1} \left[ \frac{\int u^\prime_z \, dy - \left< \int u^\prime_z \, dy \right> } {\int u^\prime_x \, dy - \left< \int u^\prime_x \, dy \right>} \right] \label{tl}
\end{equation}
The second estimation is obtained from Fourier analysis and computed from Equation (\ref{tf}), following Reference~\cite{Barkley2007mean} :
\begin{equation}
\theta_F (t) =tan^{-1}(\lambda_z/\lambda_x) \label{tf}
\end{equation}
where $\lambda=2 \pi/k$, with $k$ being the leading non-zero wavenumber identified from the power spectra (excluding the $k_x=k_z=0$ mode). The Fourier spectrum is computed for the quantity $\tau(x,z,t)$, but similar results have been observed for other observables such as $E_{cf}(x,z,t)$ and $E_v = (1/2)\int u^2_y \, dy$. The~angles can be read directly from the Fourier spectra in polar coordinates, see Figure  \ref{avis}b,d for the same values of $Re^{\scaleto{G}{4pt}}_{\tau} = 60$ and $40$, respectively. The mean angles $\widebar{\left<\theta_L\right>}$ and $\widebar{\theta}_F$ are then computed by respectively space-time-averaging and time averaging the data obtained from Equation (\ref{tl}) and (\ref{tf}).

\begin{figure}[H]
\centering
\begin{subfigure}[b]{0.48\textwidth}
\centering
\includegraphics[width=0.9\textwidth]{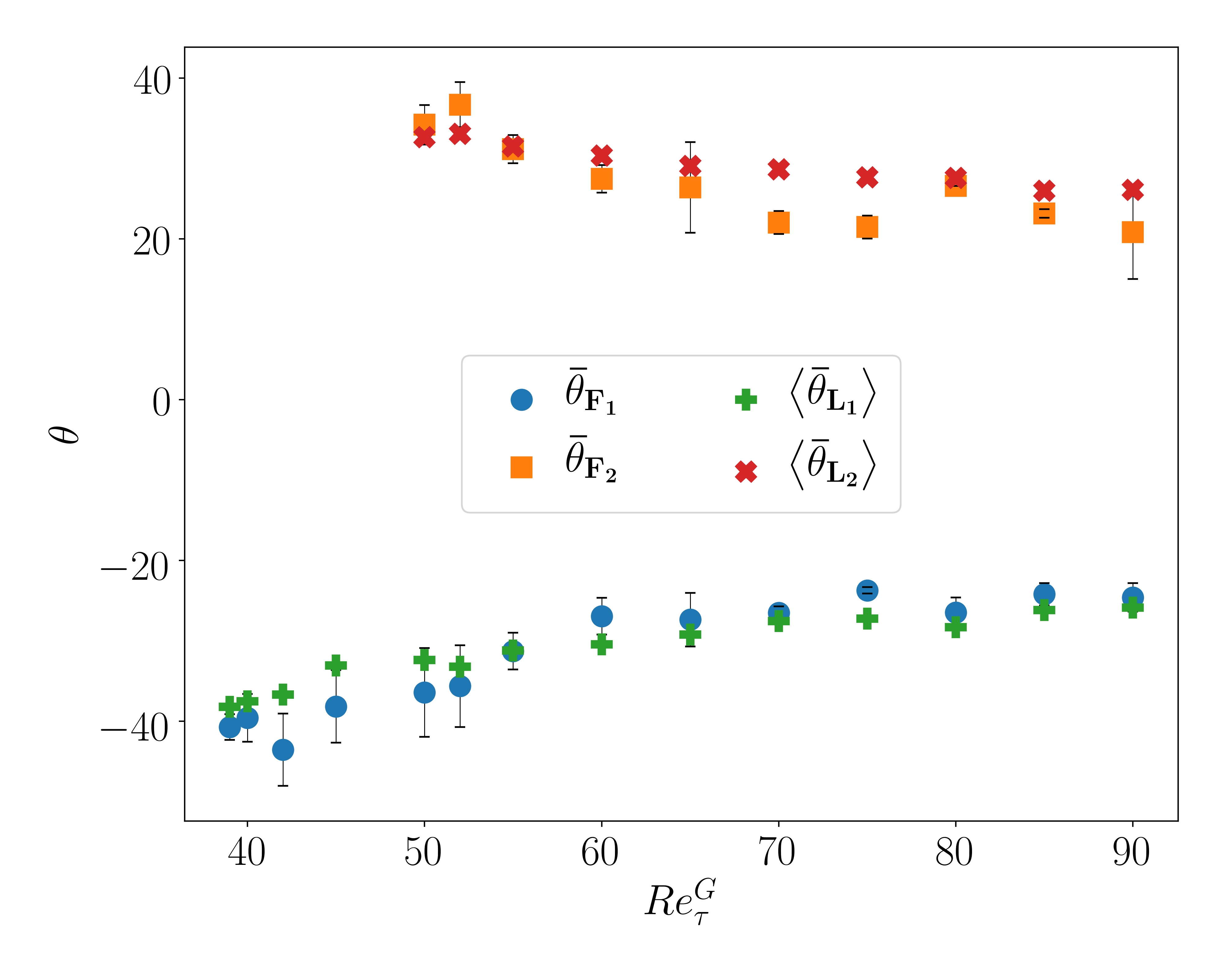}
\caption{}
\label{angle}
\end{subfigure}
\hfill
\begin{subfigure}[b]{0.48\textwidth}
\centering
\includegraphics[width=0.9\textwidth]{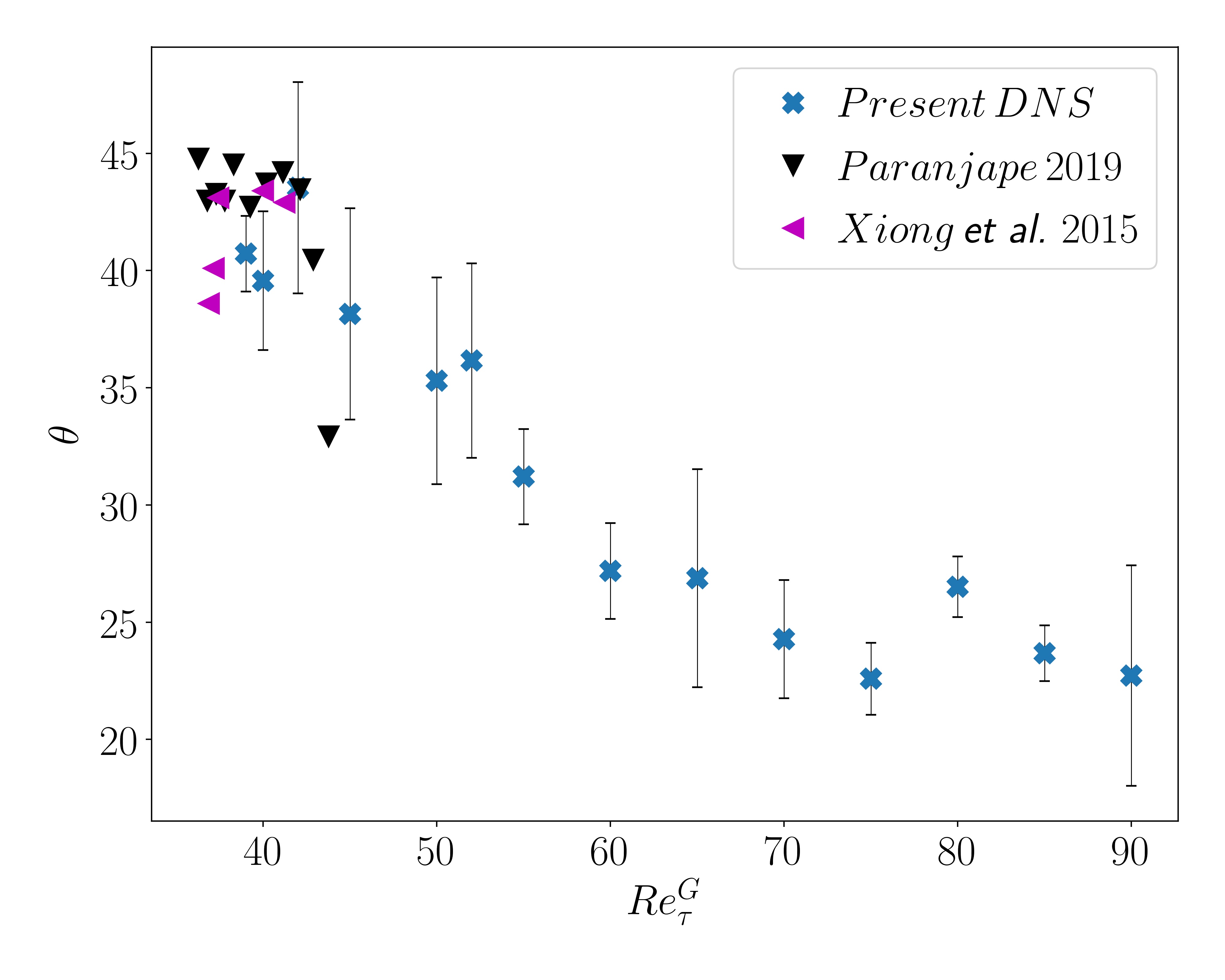}
\caption{}
\label{angle_avg}
\end{subfigure}
\caption{ ({\bf a}) Variation of the mean (signed) angle of the turbulent bands with $Re^{\scaleto{G}{4pt}}_{\tau}$, computed from the Fourier spectra ($\widebar{ \theta_{F_1}}$ , $\widebar{\theta_{F_2}}$) and the mean (signed) angle of the local velocity ($\widebar{ \left< \theta_{L_1} \right>}  $ , $\widebar{ \left< \theta _{L_2} \right>}$)  ({\bf b}) Variation of the mean unsigned band angle $\theta$ along with the data from Reference~\cite{paranjape2019thesis,Xiong2015turbulent}.}
\label{angle_plot}
\end{figure}

The variation of the mean (signed) angles with $Re^{\scaleto{G}{4pt}}_{\tau}$, computed using the two methods, is  shown in Figure \ref{angle_plot}a, where the indices $1,2$ stand for the two band orientations. Both methods provide identical results. The variation of the (unsigned) angle of the band denoted by $\theta$, computed as $\theta = \widebar{\left| \theta_F \right|}$ is shown in Figure \ref{angle_plot}b. It is found that the mean angle $\theta$ of the bands remains approximately constant with $\theta=25 ^{\circ} \pm 2.5 ^{\circ}$ in the range of values $60 \leq Re^{\scaleto{G}{4pt}}_{\tau} \leq 90$ and increases for lower value of $Re^{\scaleto{G}{4pt}}_{\tau} < 60$. In the patterning regime, i.e., for $Re^{\scaleto{G}{4pt}}_{\tau}\ge 50$, the angle of the bands is found to be distributed symmetrically with respect to zero, as a consequence of the natural symmetry $z \leftarrow -z$ of the flow. For lower $Re^{\scaleto{G}{4pt}}_{\tau}$ these quasi-regular patterns break down into individual localized structures analogous to individual puffs in cylindrical pipe flow. As the pattern dissolves, one single band orientation ends up dominating the dynamics as shown by Shimizu and Manneville \cite{shimizu2019bifurcations} for a similar domain size.
The angle $\theta$ further  increases as the regular pattern deteriorates, with $\theta_{max} \approx 40$ at $Re^{\scaleto{G}{4pt}}_{\tau}=39$.  Previous studies \cite{Tao2018extended,paranjape2019thesis} have documented that the angle of the bands approach $45 ^{\circ}$ close to the onset of transition. The present investigation agrees well with these studies (Figure \ref{angle_plot}b) while covering a wider range in Reynolds number, highlighting the difference between the puff regime for which $\theta \approx 40-45 ^{\circ}$, and the patterning regime for which $\theta$ is almost half this value (see also Figure~\ref{flowvis}).

\unskip

\begin{figure}[H]
\centering
\begin{subfigure}[b]{0.48\textwidth}
\centering
\includegraphics[width=0.9\textwidth]{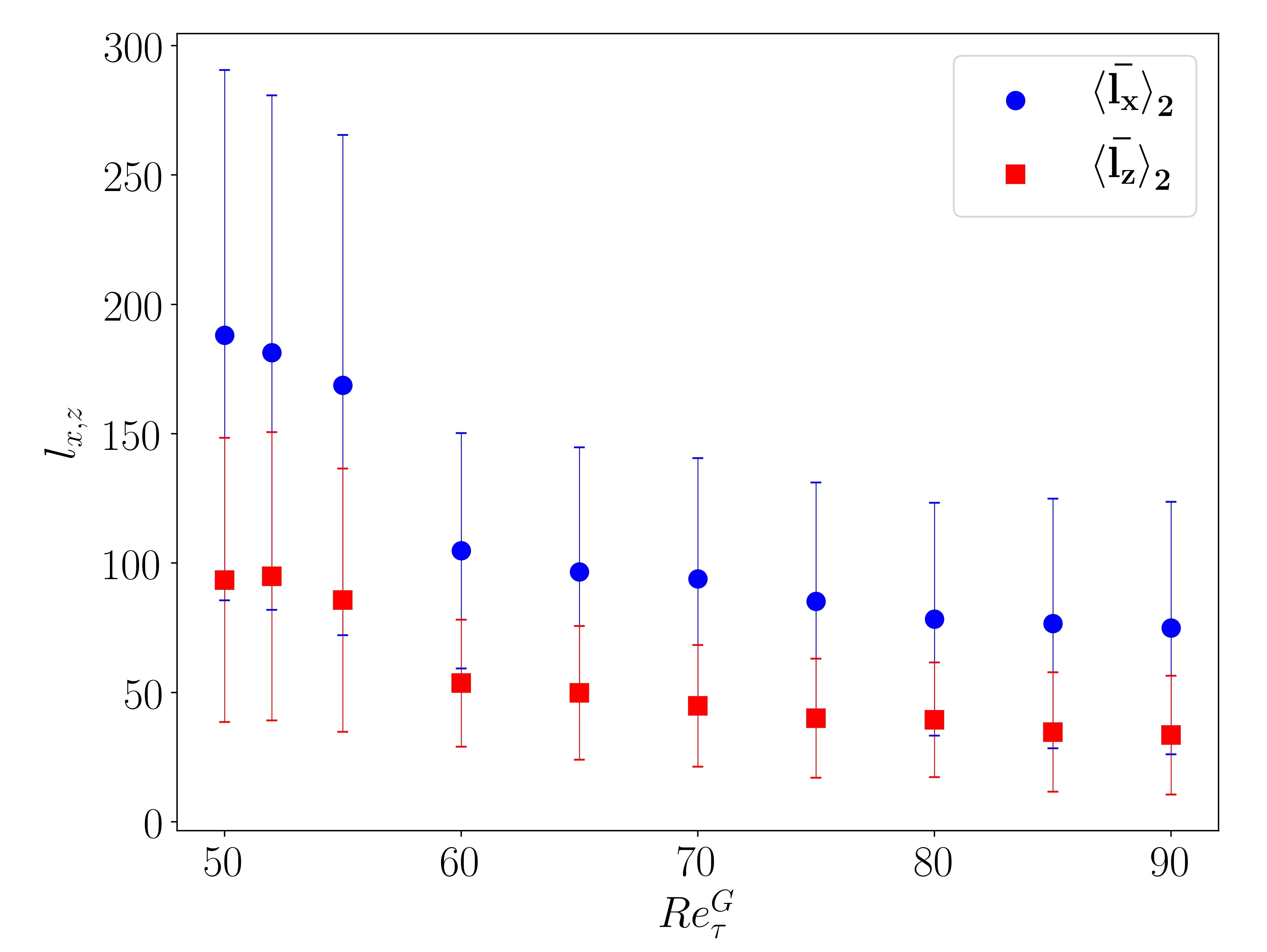}
\caption{}
\label{istd1}
\end{subfigure}
\hfill
\begin{subfigure}[b]{0.48\textwidth}
\centering
\includegraphics[width=0.9\textwidth]{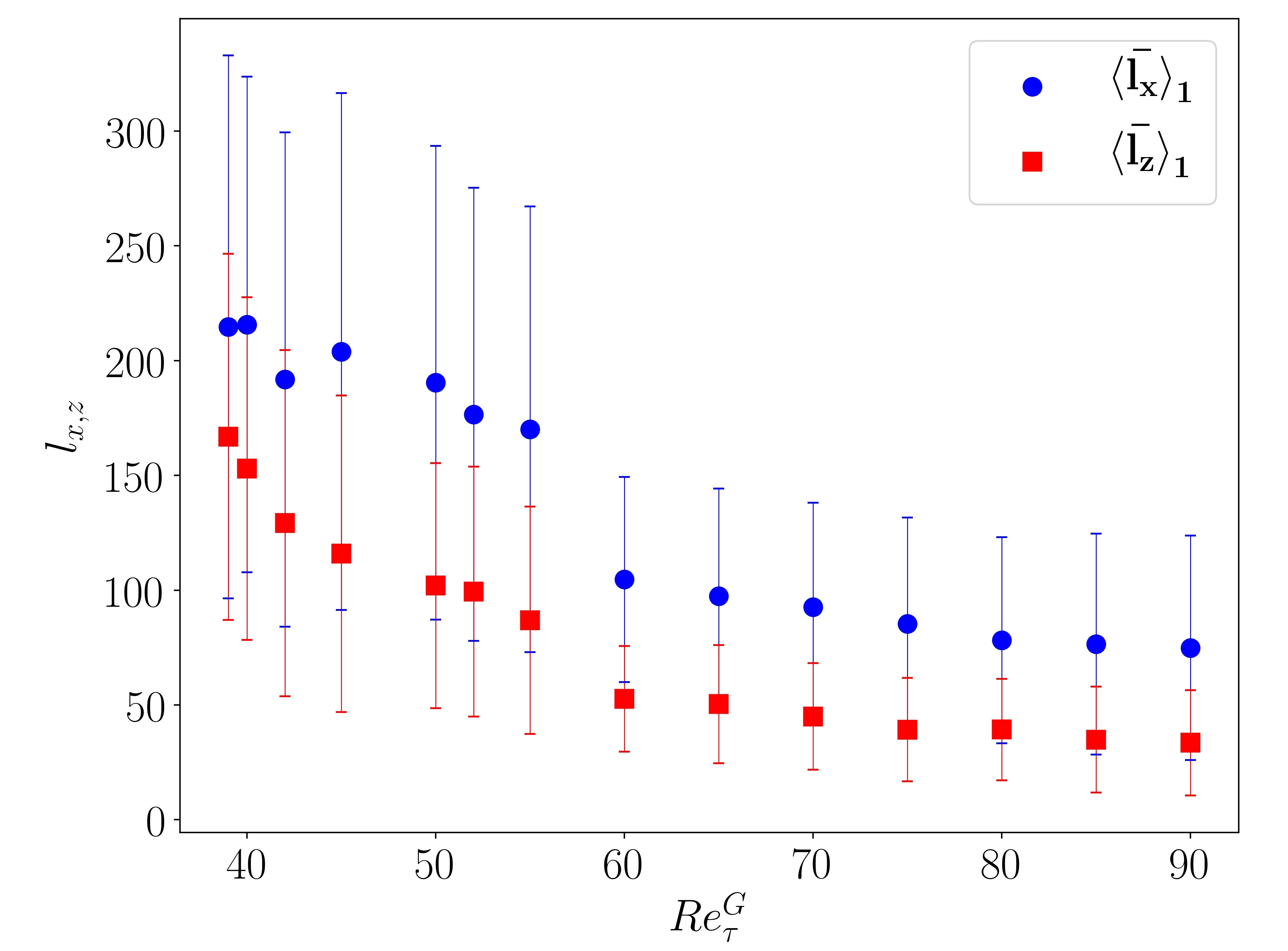}
\caption{}
\label{istd2}
\end{subfigure}
\hfill
\caption{ ({\bf a},{\bf b}) Space-time-averaged inter-stripe streamwise $\widebar{\left< l_x \right> }_{1,2}$ (blue) and spanwise $\widebar{\left< l_z \right> }_{1,2}$ (red) distances for bands of orientations $1$ and $2$, respectively.}
\label{istd}
\end{figure}

Figure~\ref{avis}c shows that across a band, the local large-scale velocity changes orientation \cite{paranjape2020oblique}. This~property is used to sort out the local maxima of $\tau$ (higher than $\left< \tau \right> + \sigma(\tau)$) as belonging to one band with a particular inclination. This allows {one} to define the respective streamwise and spanwise inter-stripe distances $l_x$ and $l_z$ between bands of the same orientation. Figure~\ref{istd}a,b displays $\widebar{\left<l_x\right>} \,  and \, \widebar{\left<l_z\right>} $ for orientations $1$ and $2$, respectively, as a function of $Re^{\scaleto{G}{4pt}}_{\tau}$. Both increase when decreasing $Re^{\scaleto{G}{4pt}}_{\tau}$. They vary in parallel in the patterning regime, hence the quasi-constant angle $\theta$ of the bands. When only one band orientation survives, one observes that the increase in $\theta$ amounts to the saturation of $\widebar{\left<l_x\right>}_{1} $, while $\widebar{\left<l_z\right>}_{1} $ keeps increasing.

%

%

\subsection{Global Variables: Moody Diagram }
\label{scf}

\begin{figure}[H]
\centering
\includegraphics[width=0.65\textwidth]{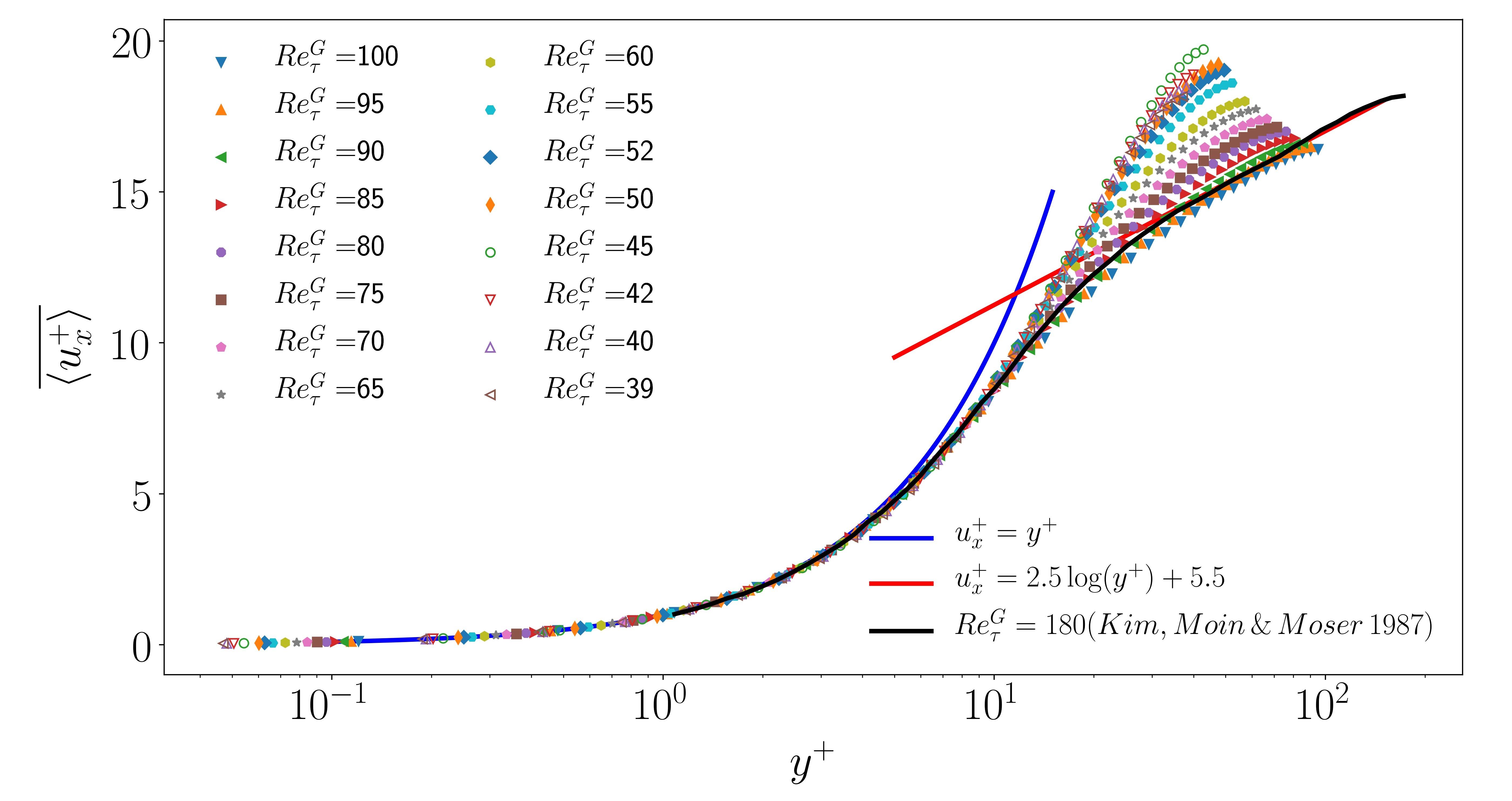}

\caption{Mean flow profile $u^{+}_x (y^{+})$ for $Re^G_\tau$ from 100 down to 39. Blue: law of the wall $u^{+}_x=y^{+}$, red: logarithmic law of the wall $u^{+}_x=2.5 \, \log(y^{+}) + 5.5$, black: DNS by Kim, Moin and Moser from  Reference~\cite{kim_turbulence_1987}.}
\label{Umean}
\end{figure}

{The mean velocity profile $\widebar{\left<u^{+}_x\right>}$ is defined
as the average of $u_x$ over $x$,$z$ and $t$, expressed in units
of $u_{\tau}$. It is shown in Figure \ref{Umean} as a function of $y^+=yu_{\tau}/\nu$
and compared with the classical DNS data by Kim  {et al.} \cite{kim_turbulence_1987}
obtained at higher $Re_{\tau}^G=180$. The whole figure is similar to figures 3 and 10 in Reference \cite{Tsukahara2005dns,Tsukahara2010transition}, respectively.
As expected for the present low values of $Re_{\tau}^G$, the velocity field matches the linearized profile $u_x^+=y^+$
next to the wall but does not develop a logarithmic dependence with respect to $y^+$.} \\

\begin{figure}[H]
\centering
\includegraphics[width=0.5\textwidth]{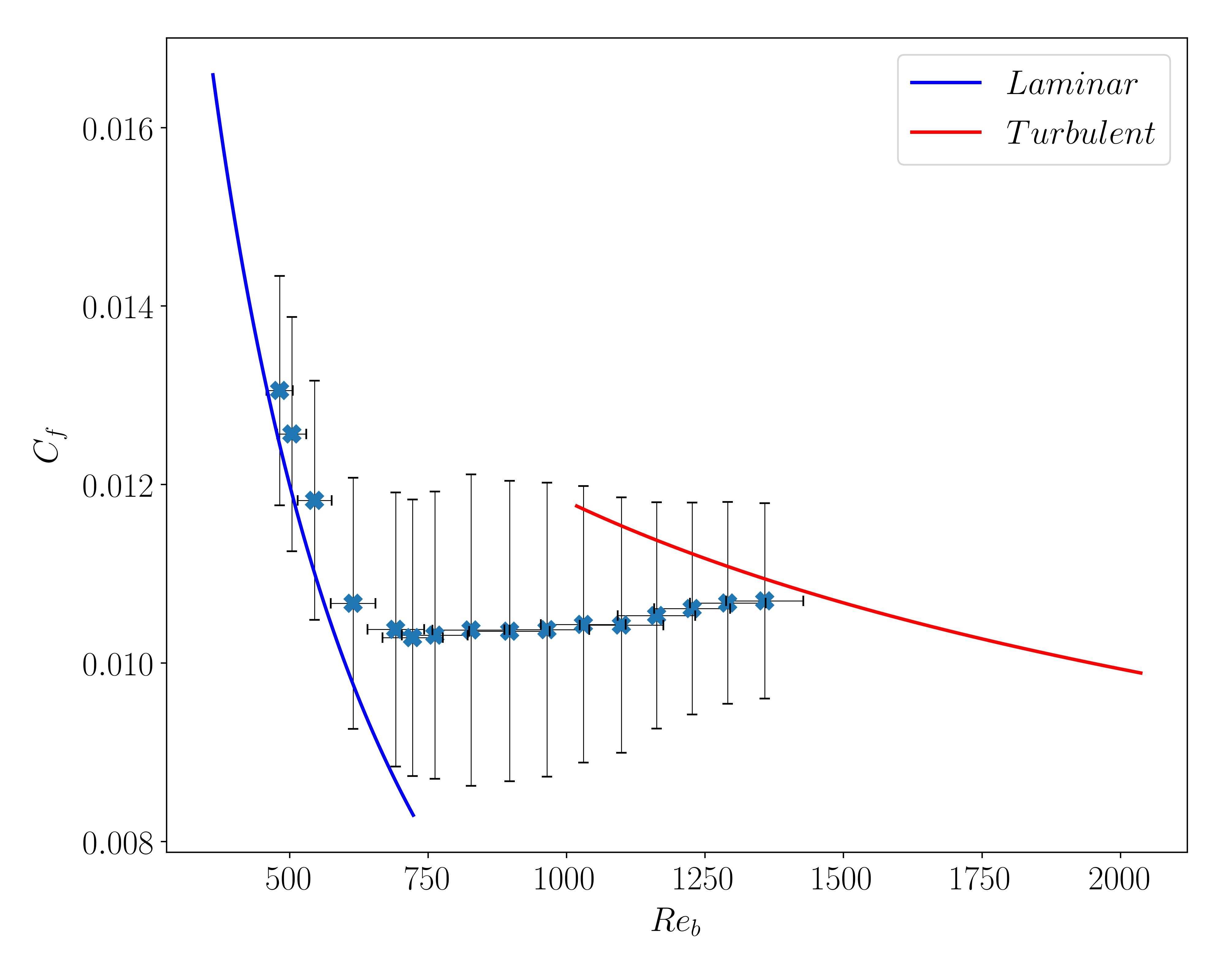}
\caption{Friction coefficient $C_f$ vs. $Re_b$, with horizontal and vertical error bars indicating the fluctuations these quantities would inherit from that of the field $u_b$ (see text for details).}
\label{cf}
\end{figure}

At a global level of description, the laminar and turbulent flow are traditionally represented in the classical Moody diagram in which the Fanning friction factor $C_f$ defined as the ratio between the pressure drop along the channel length and the kinetic energy per unit volume based on the mean bulk velocity $U_b = \widebar{\left<u_b\right>}$,
\begin{equation}
C_f =  \frac{\left| \Delta p \right|}{1/2 \, \rho U_b^2 } \, \frac{h}{L_x} = \frac{ \widebar{ \left< \tau \right> } } {1/2 \, \rho U_b^2} = \frac{2 \, {Re^{\scaleto{G}{4pt}}_{\tau}}^2}{Re_b^2}, \label{cfe}
\end{equation}
is traditionally plotted versus $Re_b$ as shown with plain symbols in Figure \ref{cf}.
{Another way to express $C_f$ is to use inner units, in which case $C_f=2/(u_b^+)^2$, with $u_b^+=u_b/u_{\tau}$. $C_f$ is then linked only to the integral of the mean profile displayed in Figure \ref{Umean}.}

For the laminar flow, the dependence of $C_f$ vs. $Re_b$ is analytically given by $Cf_{lam}=6/Re_b$ (blue continuous line). In the featureless turbulent regime, it is known empirically as the Blasius' friction law scaling $\widebar{Re_b}^{-1/4}$ (red continuous line).
For intermediate values of $Re_b$,  $C_f$ clearly deviates from the turbulent branch, and remains far from the laminar value \cite{dean1978reynolds}. Here we notice, in agreement with \cite{Xiong2015turbulent} and \cite{shimizu2019bifurcations} that $C_f \approx 0.01$ remains essentially constant in this transitional regime. What is remarkable is that this regime of constant $C_f$ coincides with the patterning regime observed for $50 \leq Re^{\scaleto{G}{4pt}}_{\tau} \leq 90$, corresponding to $690 \leq Re_b \leq 1225 $, \emph{as if the respective amount of turbulent and laminar domains was precisely ensuring $C_f= cst$}. As the pattern fractures, $C_f$ increases and approaches the laminar curve. We note that the observation of this property requires large computational domains to be observed, which explains why it had not been noticed until recently, even in experiments.

Given the complex spatio-temporal dynamics in the transitional regime, the bulk velocity $u_b$ is expected to strongly fluctuate both in space and time. We also report in Figure \ref{cf}, how these fluctuations would translate on $Re_b$ and $C_f$, if the latter were computed using the locally fluctuating field $u_b$ instead of its mean value $U_b$. These fluctuations are significant (up to 10--15\%) and suggest to further explore them, which is the topic of the next section and the main focus of the present work.

\subsection{Joint Probability Distribution of $Re_\tau$ and $Re_b$}\label{sjpdf}

\textls[-15]{Reynolds numbers such as $Re_\tau$ and $Re_b$ are traditionally seen as global parameters characterizing} the flow.~They are defined based on velocity scales obtained from space-time average. It~is~straightforward to extend these definitions to the local fields $Re_b(x,z,t)~=~u_b(x,z,t) h/\nu$ and $Re_{\tau} = u_{\tau}(x,z,t) h / \nu$, with $u_{\tau}(x,z,t)= (\tau(x,z,t)/\rho)^{1/2}$. Please note that with this definition, $\widebar{\left<Re_{\tau}\right>}$ is not strictly equal to the imposed $Re^{\scaleto{G}{4pt}}_{\tau}$, because of the nonlinear relation between $Re_{\tau}$ and $\tau$.

\begin{figure}[H]
\centering
\scalebox{0.85}{\begin{tikzpicture}
\node[inner sep=0pt] (f1) at (0,0)
{\hypertarget{fig9a3}{\includegraphics[scale=0.25,trim=0.8cm 1.8cm 3.6cm 3.6cm,clip]{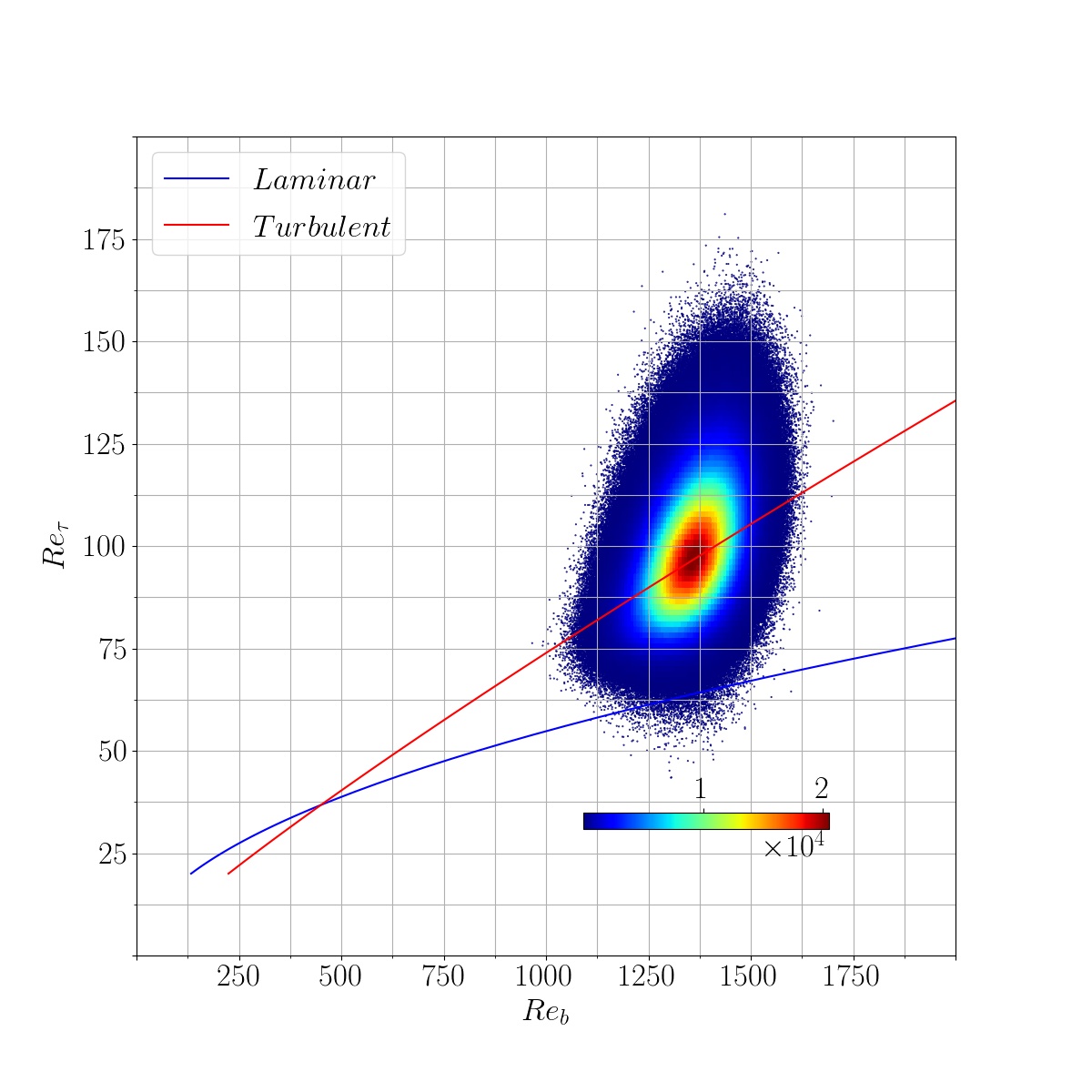}}};
\node[inner sep=0pt] (f2) at (4.3,0)
{\includegraphics[scale=0.25,trim=3.8cm 1.8cm 3.6cm 3.6cm,clip]{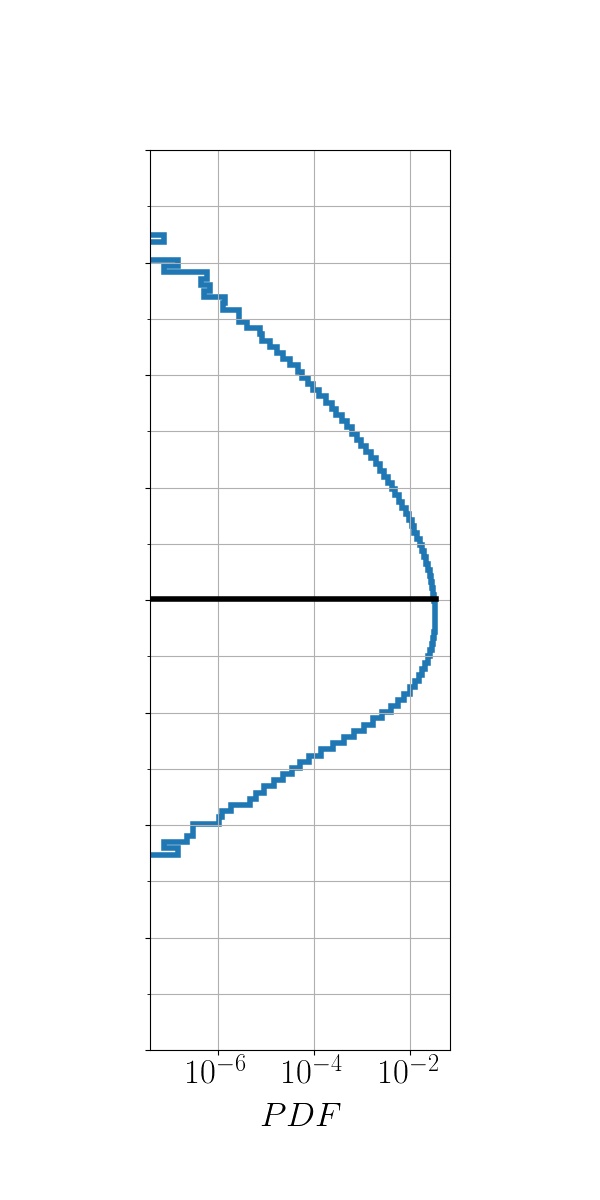}};
\node[inner sep=0pt] (f3) at (0,4.2)
{\includegraphics[scale=0.25,trim=0.8cm 3.8cm 3.6cm 3.4cm,clip]{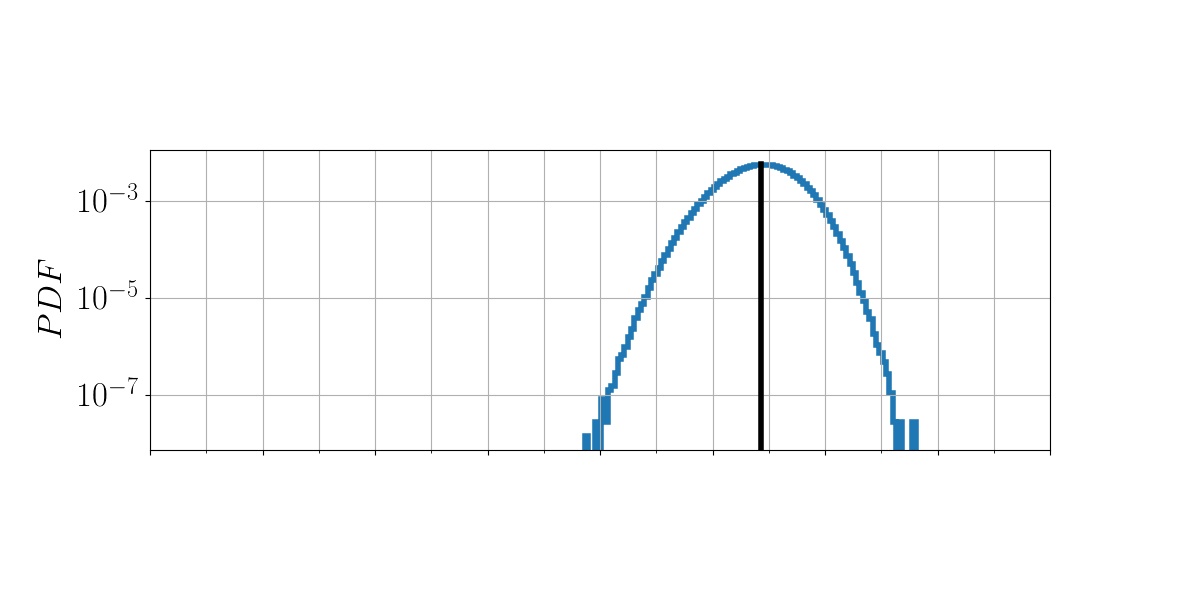}};
\node[inner sep=0pt] at (-1.5,1) {\fontsize{10pt}{10pt} $\mathbf{(a_1)}$ };
\node[inner sep=0pt] at (-1.5,4.5) {\fontsize{10pt}{10pt} $\mathbf{(a_2)}$ };
\node[inner sep=0pt] at (4.5,2.5) {\fontsize{10pt}{10pt} $\mathbf{(a_3)}$ };

\node[inner sep=0pt] (f1) at (8.6,0)
{\includegraphics[scale=0.25,trim=0.8cm 1.8cm 3.6cm 3.6cm,clip]{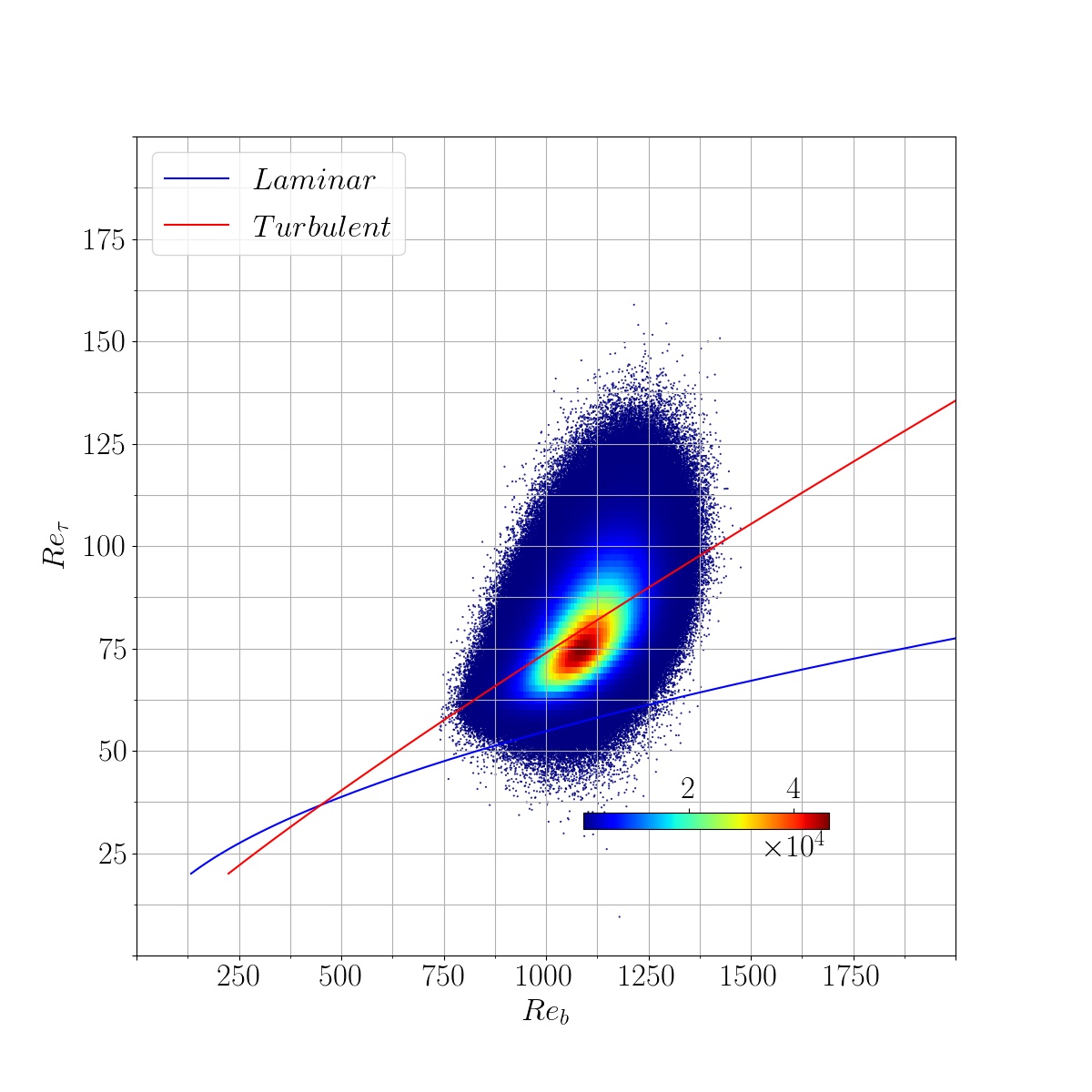}};
\node[inner sep=0pt] (f2) at (12.9,0)
{\includegraphics[scale=0.25,trim=3.8cm 1.8cm 3.6cm 3.6cm,clip]{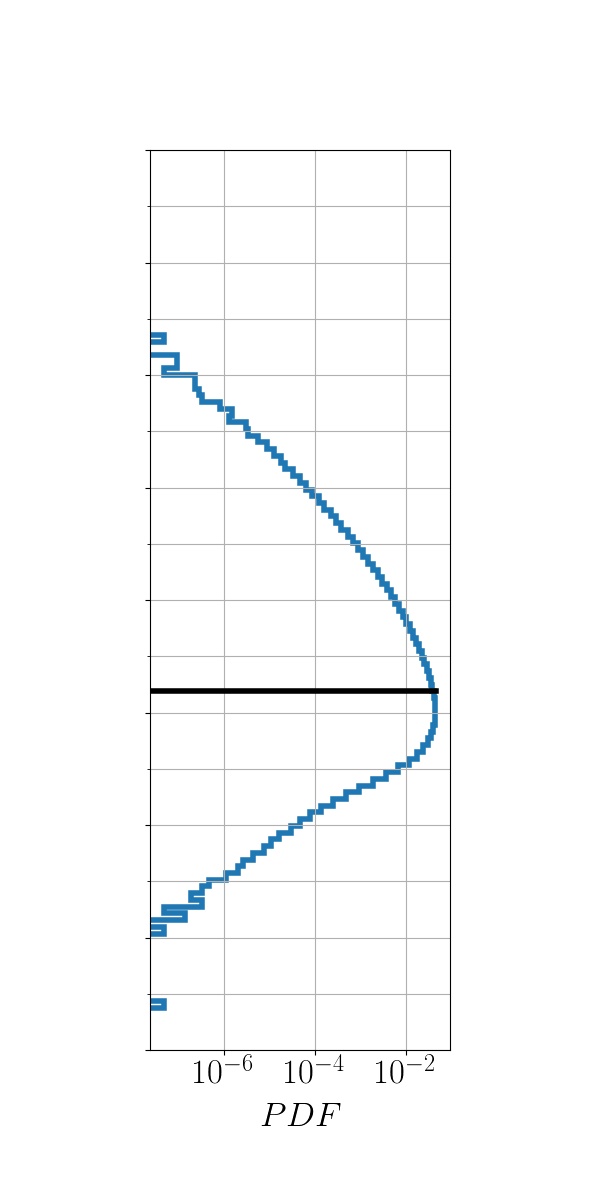}};
\node[inner sep=0pt] (f3) at (8.6,4.2)
{\includegraphics[scale=0.25,trim=0.8cm 3.8cm 3.6cm 3.6cm,clip]{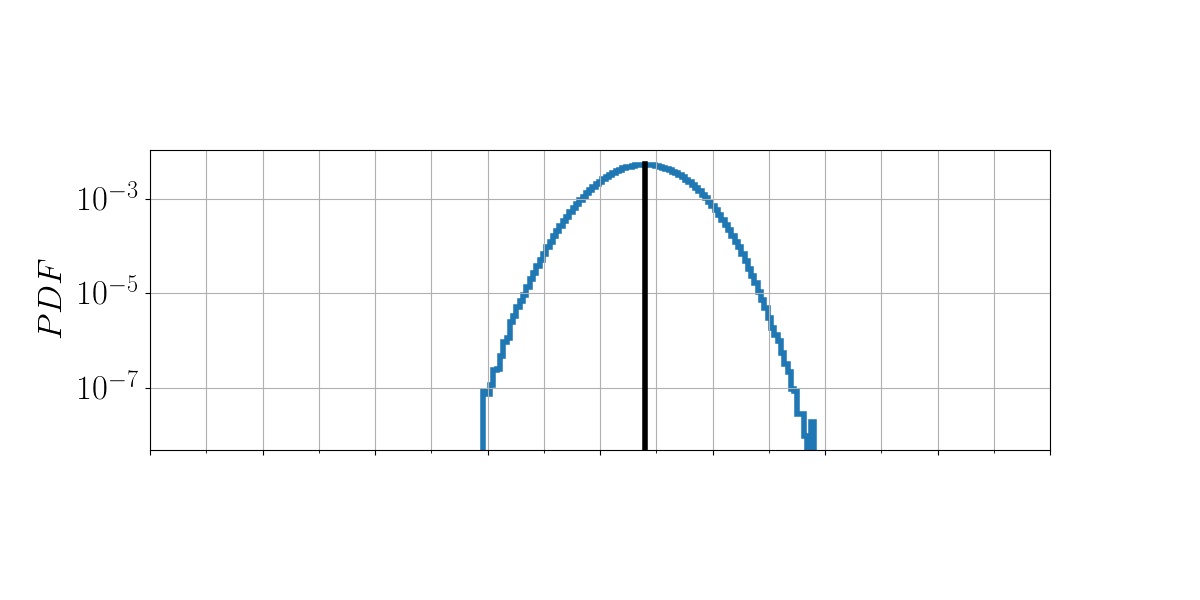}};
\node[inner sep=0pt] at (7.1,1) {\fontsize{10pt}{10pt} $\mathbf{(b_1)}$ };
\node[inner sep=0pt] at (7.1,4.5) {\fontsize{10pt}{10pt} $\mathbf{(b_2)}$ };
\node[inner sep=0pt] at (13.1,2.5) {\fontsize{10pt}{10pt} $\mathbf{(b_3)}$ };

\node[inner sep=0pt] (f1) at (0,-9.2)
{\includegraphics[scale=0.25,trim=0.8cm 1.8cm 3.6cm 3.6cm,clip]{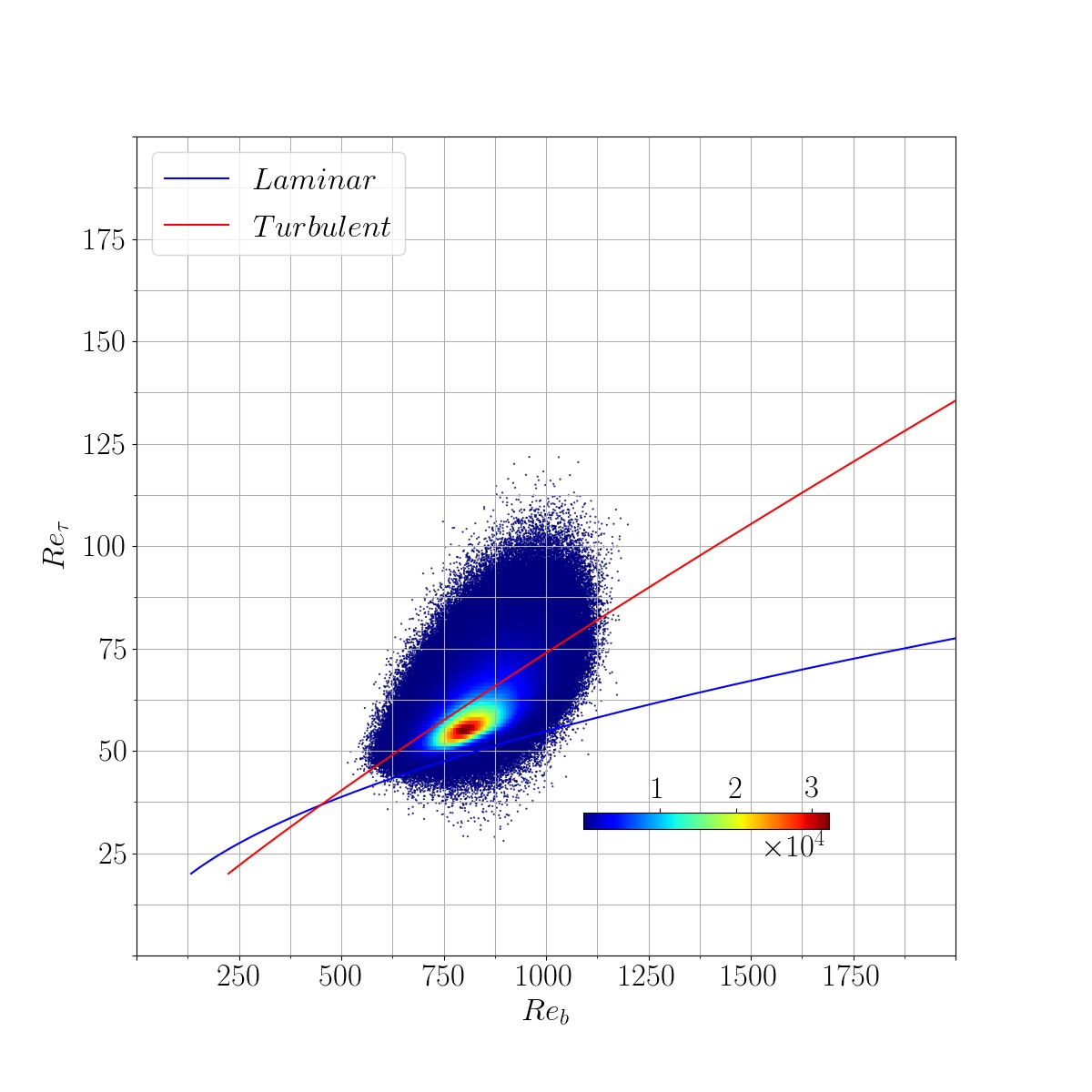}};
\node[inner sep=0pt] (f2) at (4.3,-9.2)
{\includegraphics[scale=0.25,trim=3.8cm 1.8cm 3.6cm 3.6cm,clip]{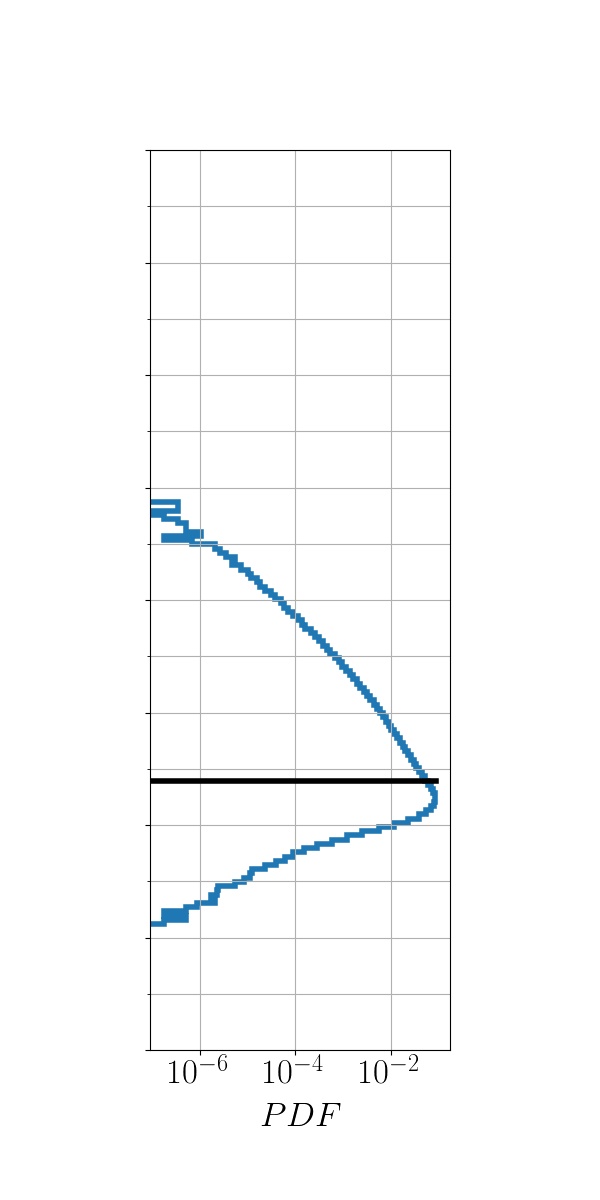}};
\node[inner sep=0pt] (f3) at (0,-5)
{\includegraphics[scale=0.25,trim=0.8cm 3.8cm 3.6cm 3.4cm,clip]{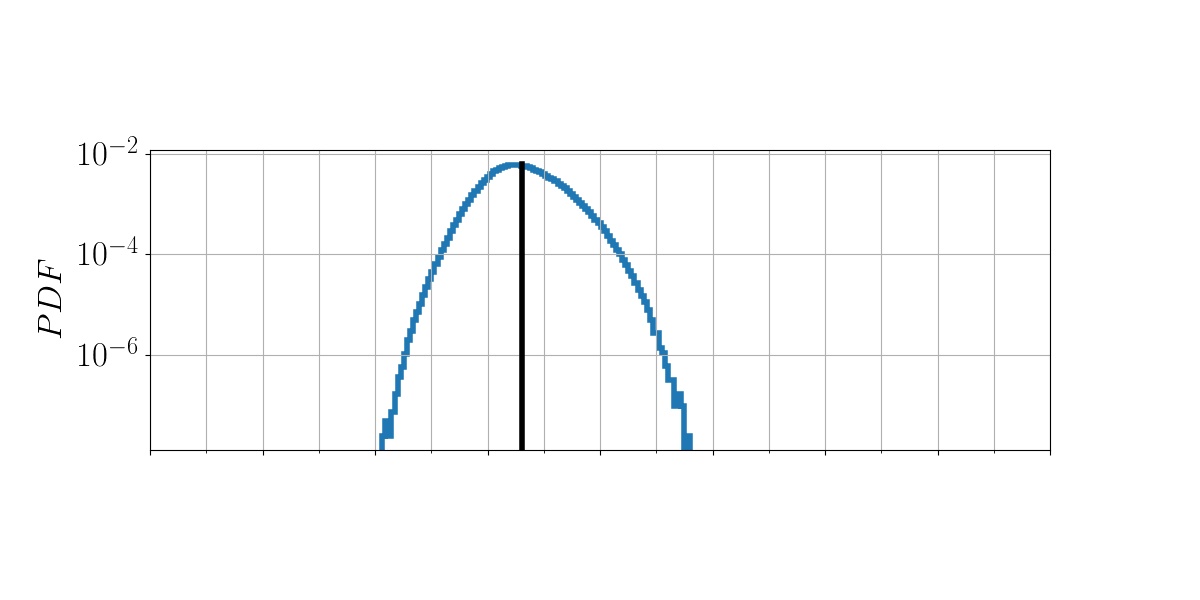}};
\node[inner sep=0pt] at (-1.5,-8) {\fontsize{10pt}{10pt} $\mathbf{(c_1)}$ };
\node[inner sep=0pt] at (-2,-5) {\fontsize{10pt}{10pt} $\mathbf{(c_2)}$ };
\node[inner sep=0pt] at (4.5,-7.5) {\fontsize{10pt}{10pt} $\mathbf{(c_3)}$ };

\node[inner sep=0pt] (f1) at (8.6,-9.2)
{\includegraphics[scale=0.25,trim=0.8cm 1.8cm 3.6cm 3.6cm,clip]{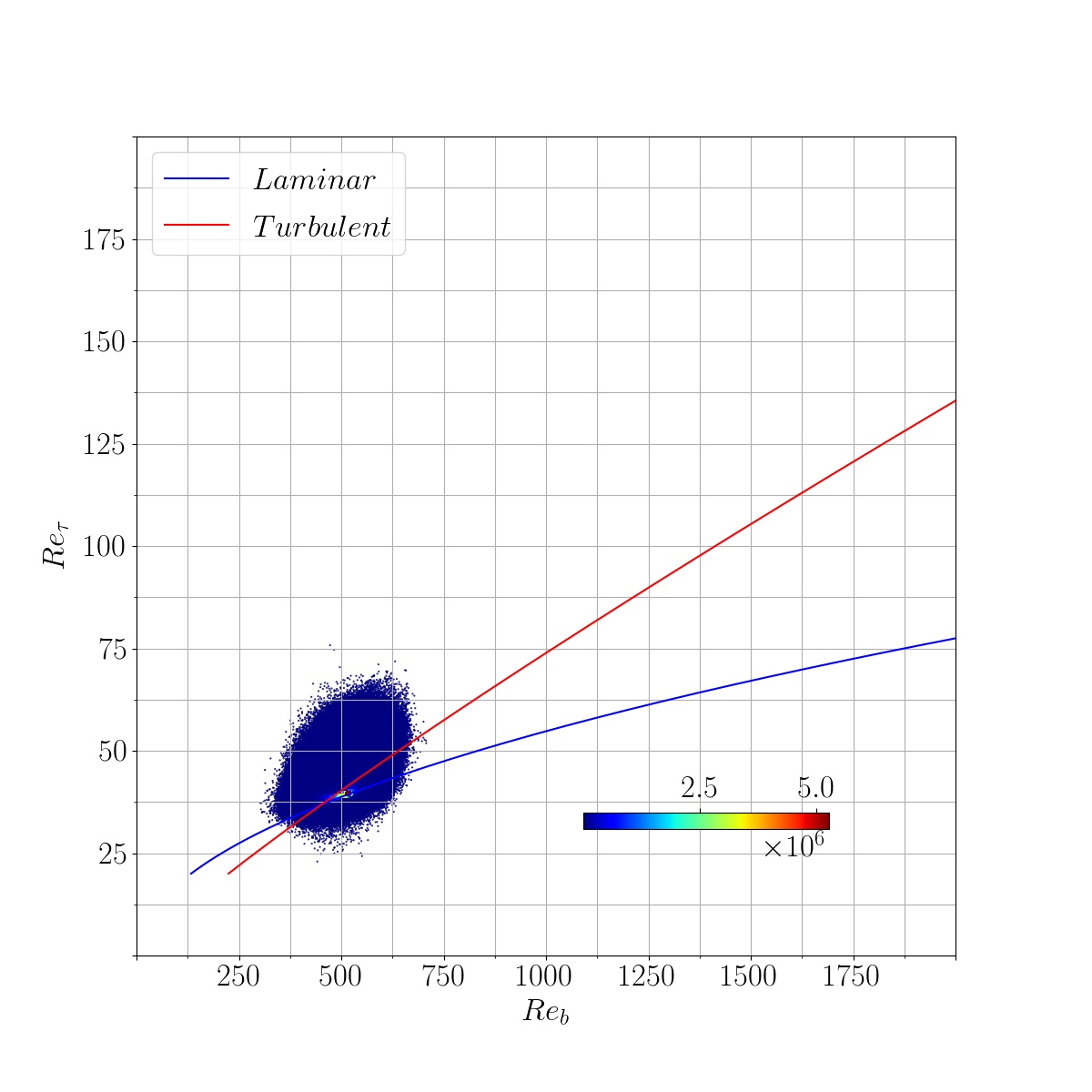}};
\node[inner sep=0pt] (f2) at (12.9,-9.2)
{\includegraphics[scale=0.25,trim=3.8cm 1.8cm 3.6cm 3.6cm,clip]{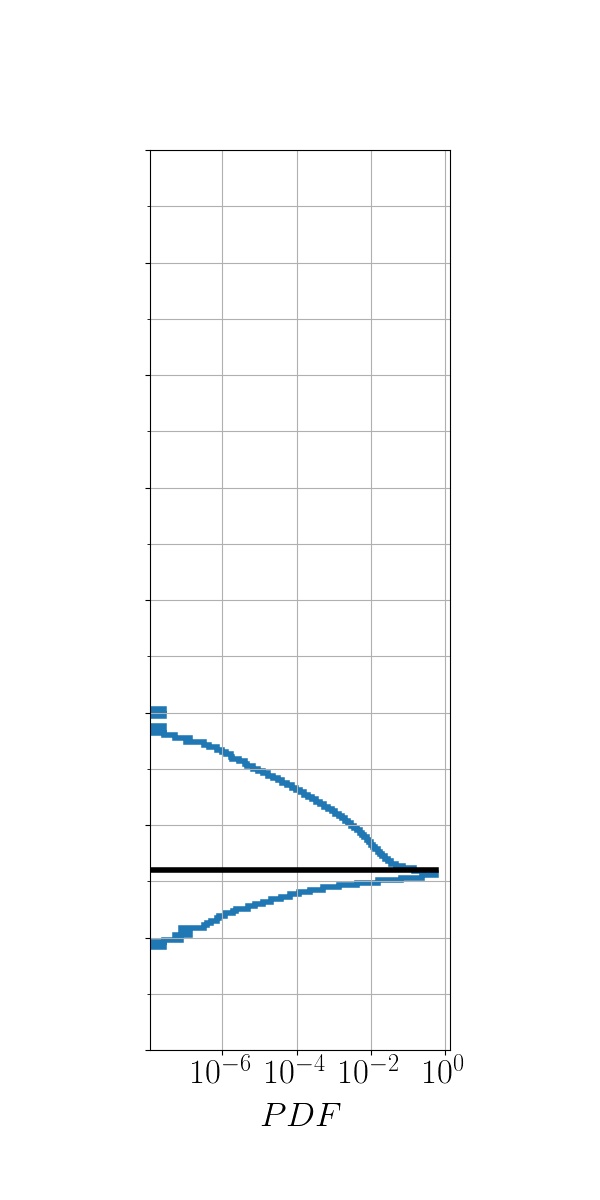}};
\node[inner sep=0pt] (f3) at (8.6,-5)
{\includegraphics[scale=0.25,trim=0.8cm 3.8cm 3.6cm 3.6cm,clip]{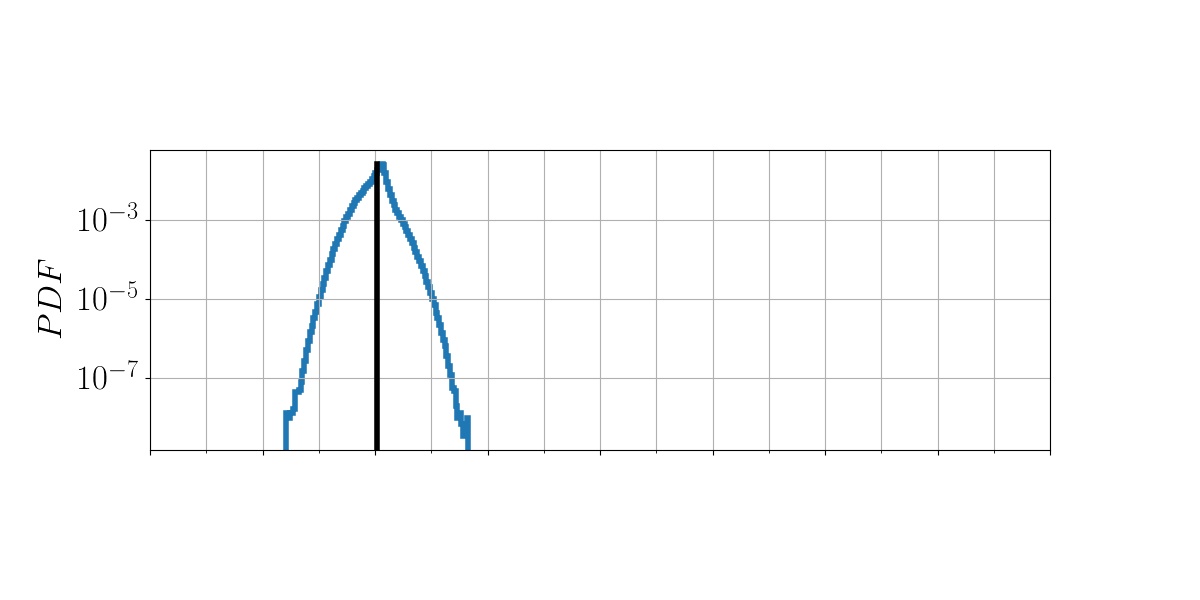}};
\node[inner sep=0pt] at (7.1,-8) {\fontsize{10pt}{10pt} $\mathbf{(d_1)}$ };
\node[inner sep=0pt] at (9.5,-4.5) {\fontsize{10pt}{10pt} $\mathbf{(d_2)}$ };
\node[inner sep=0pt] at (13.1,-7.5) {\fontsize{10pt}{10pt} $\mathbf{(d_3)}$ };

\end{tikzpicture} }

\caption{\textls[-5]{$\mathbf{(a_1) \, (b_1) \, (c_1) \, (d_1)}$ Joint probability distribution of the quantities $Re_b$ and $Re_{\tau}$ for $Re^{\scaleto{G}{4pt}}_{\tau}=100,~80,$} 60, 40 together with their marginal distribution shown in lin-log scale for $Re_b$ in $\mathbf{(a_2) \, (b_2) \, (c_2) \, (d_2)}$ and for $Re_{\tau}$ in $\mathbf{(a_3) \, (b_3) \, (c_3) \, (d_3)}$ with the mean value indicated by a vertical/horizontal black line.}
\label{pdf}
\end{figure}
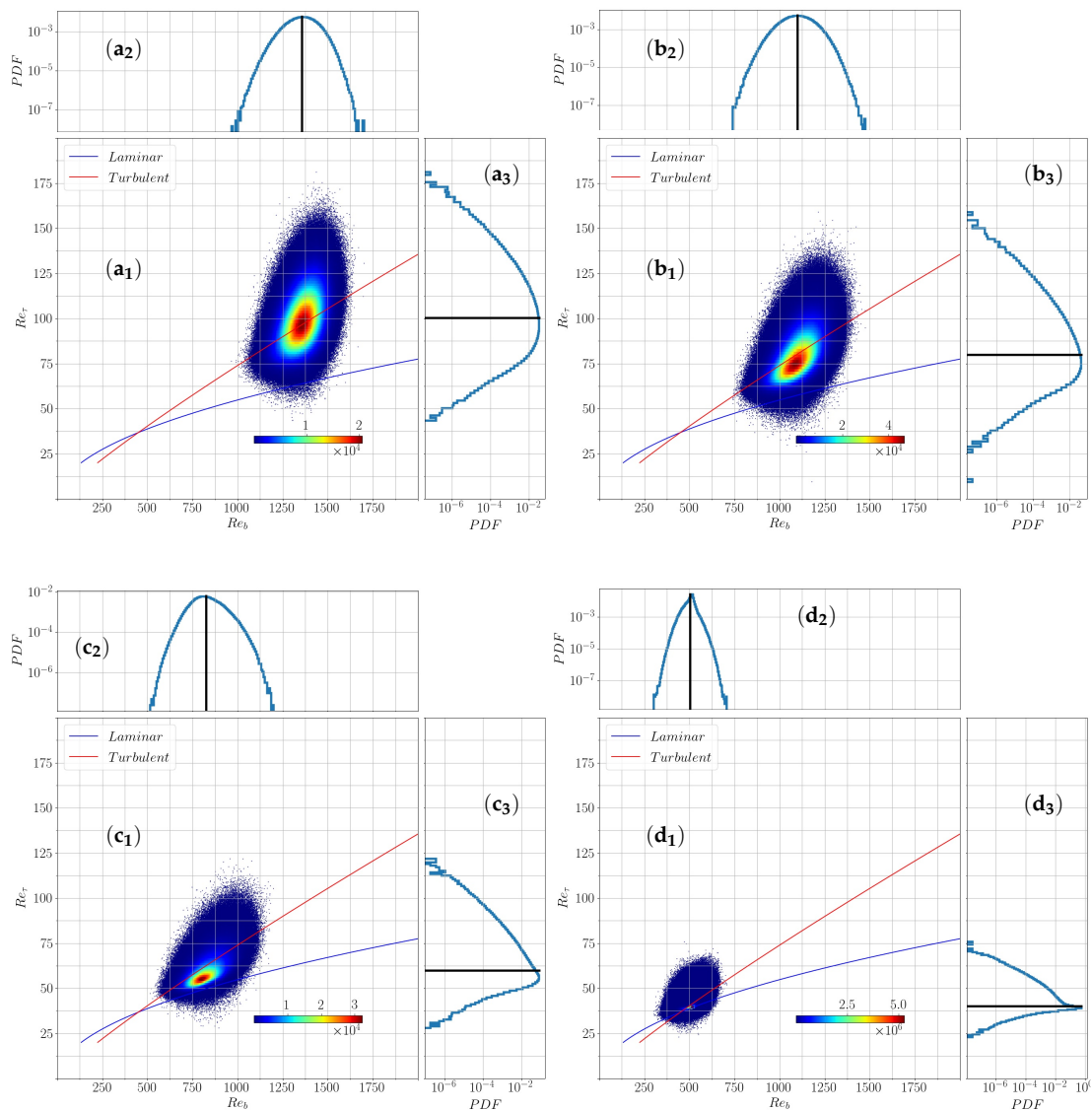

Investigation of the entire transitional regime is provided through a two-dimensional state portrait ($Re_b - Re_{\tau}$) constructed from this local definition of the Reynolds number. The joint probability density distribution is constructed in this state space with the space-time data for different $Re^{\scaleto{G}{4pt}}_{\tau}$. The state space for $Re^{\scaleto{G}{4pt}}_{\tau}=100,80,60,40$ is shown in Figure \ref{pdf}. The continuous blue and red lines again correspond to the scalings known analytically {for the laminar flow, and empirically for featureless turbulent flows for high enough Reynolds numbers}. As expected the most probable values of $Re_b$ and $Re_{\tau}$, follow the same trend as their global counterpart: they match the continuous curve in the featureless turbulent regime, and progressively depart from it
to move towards the laminar branch at the lowest $Re^{\scaleto{G}{4pt}}_{\tau}$ explored here. More interesting are the distributions. First, we observe that the relative fluctuations are significantly larger for $Re_{\tau}$ than for $Re_b$, the difference being larger for the larger $Re^{\scaleto{G}{4pt}}_{\tau}$.
Secondly the distributions are not simple Gaussians. Even in the featureless turbulent regime, the marginal distribution of $Re_{\tau}$ is already relatively skewed (Figure~\ref{pdf}{(a$_3$)}).

As $Re^{\scaleto{G}{4pt}}_{\tau}$ is reduced, the overall width of the distribution decreases, but the shape of the marginal distributions of $Re_{\tau}$ differs more and more from a Gaussian. More specifically, although the distribution remains unimodal, we note that the marginal distribution of $Re_{\tau}$ is more and more skewed. We~also note that the right wing of the distribution is not convex anymore.~To further quantify these observations, a systematic analysis of the moments of this distribution is conducted in the next~section.

\subsection{Higher-Order Statistics} \label{stm}

The higher-order statistics of $Re_\tau$, $Re_b$ and $E_{cf}$ are presented in this section. For any field $A~=~A(x,z,t)$, we compute the spatio-temporal average $m=\widebar{\left<A\right>}$, the variance $\sigma ^2= \widebar{\left<(A-m)^2\right>}$ and the $k^{th}$ standardized higher-order moment $\widebar{\left<(A-m)^k\right>}/\sigma^k$ (for $k \ge 3$).

Their mean values of $Re_b$ and $Re_{\tau}$ (Figure \ref{stat}a) simply follow the trends described above for the most probable value of the distribution, connecting the turbulent and the laminar branch, when $Re^{\scaleto{G}{4pt}}_{\tau}$ decreases.
Away from the turbulent and laminar branches $Re_{\tau}$ is linearly related to $Re_b$, in agreement with the observation of a constant $C_f$. The standard deviation $\sigma$ (Figure \ref{stat}b) for $Re_{\tau}$ and $Re_b$ decrease together with $Re^{\scaleto{G}{4pt}}_{\tau}$. This decreasing trend agrees well with the experimental wall shear stress data reported in Reference~\cite{agrawal2020investigating}. The standard deviation for $E_{cf}$ is found to increase with decreasing $Re^{\scaleto{G}{4pt}}_{\tau}$, matching the trend reported in Reference~\cite{shimizu2019bifurcations}.

The variation of the 3{rd} and 4{th} moments $m_3$ and $m_4$, i.e., the Skewness ($S$) and Kurtosis ($K$), versus $Re^{\scaleto{G}{4pt}}_{\tau}$ for the observable $Re_{\tau}$ and $E_{cf}$ is shown in Figure \ref{stat}c. These moments exhibit a strongly increasing trend with reducing $Re^{\scaleto{G}{4pt}}_{\tau}$ for both quantities. This similarity in behavior leads to $K \propto S^2$ as shown in Figure \ref{stat}e. This correlation between the third and fourth statistical moments was first noted in Reference~\cite{jovanovic_statistical_1993} for the fluctuating velocity in turbulent boundary layers at high Reynolds number. In~the~transitional regime, the same relationship has been found to hold in the experiments of \mbox{Agrawal {et al.} \cite{agrawal2020investigating}} from wall shear stress data. We therefore confirm this yet-to-be-understood extension of a high Reynolds number scaling down to the spatio-temporal intermittent regime. Furthermore, we observe that the same scaling also holds for the turbulent kinetic energy $E_{cf}$ (Figure~\ref{stat}e). In contrast it does not apply to $Re_b$ (inset of Figure~\ref{stat}e).~The reason is that while the Kurtosis follows the same trend as for the two other observables, (Figure \ref{stat}d), the skewness shows a markedly different behavior: it~is~non-monotonous, changes sign twice and exhibit a maximum in the core of the spatio-temporal intermittent regime.

\begin{figure}[H]
\centering
\begin{subfigure}[b]{0.45\textwidth}
\centering
\includegraphics[width=\textwidth]{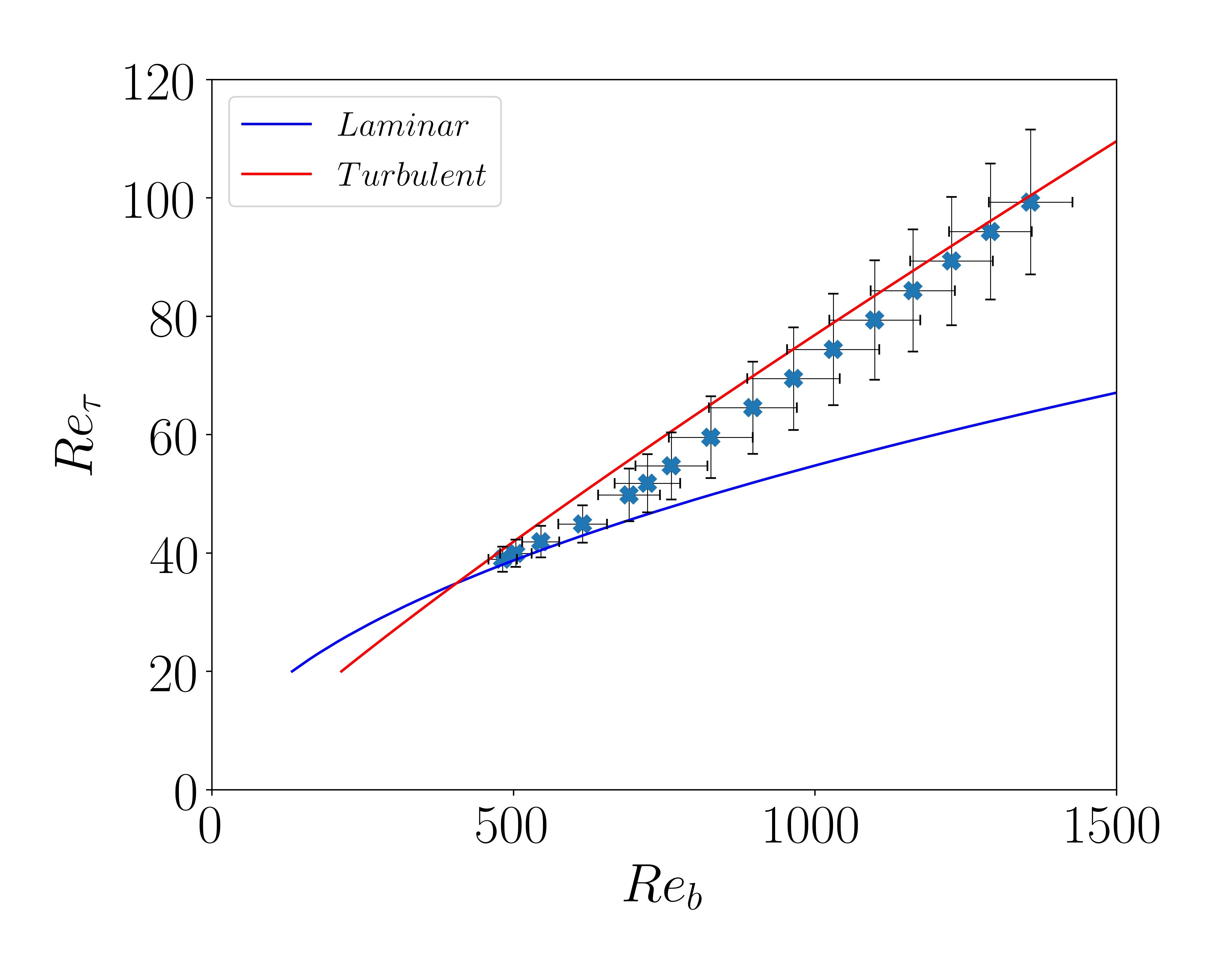}
\caption{}
\label{st1}
\end{subfigure}
\hfill
\begin{subfigure}[b]{0.45\textwidth}
\centering
\includegraphics[width=\textwidth]{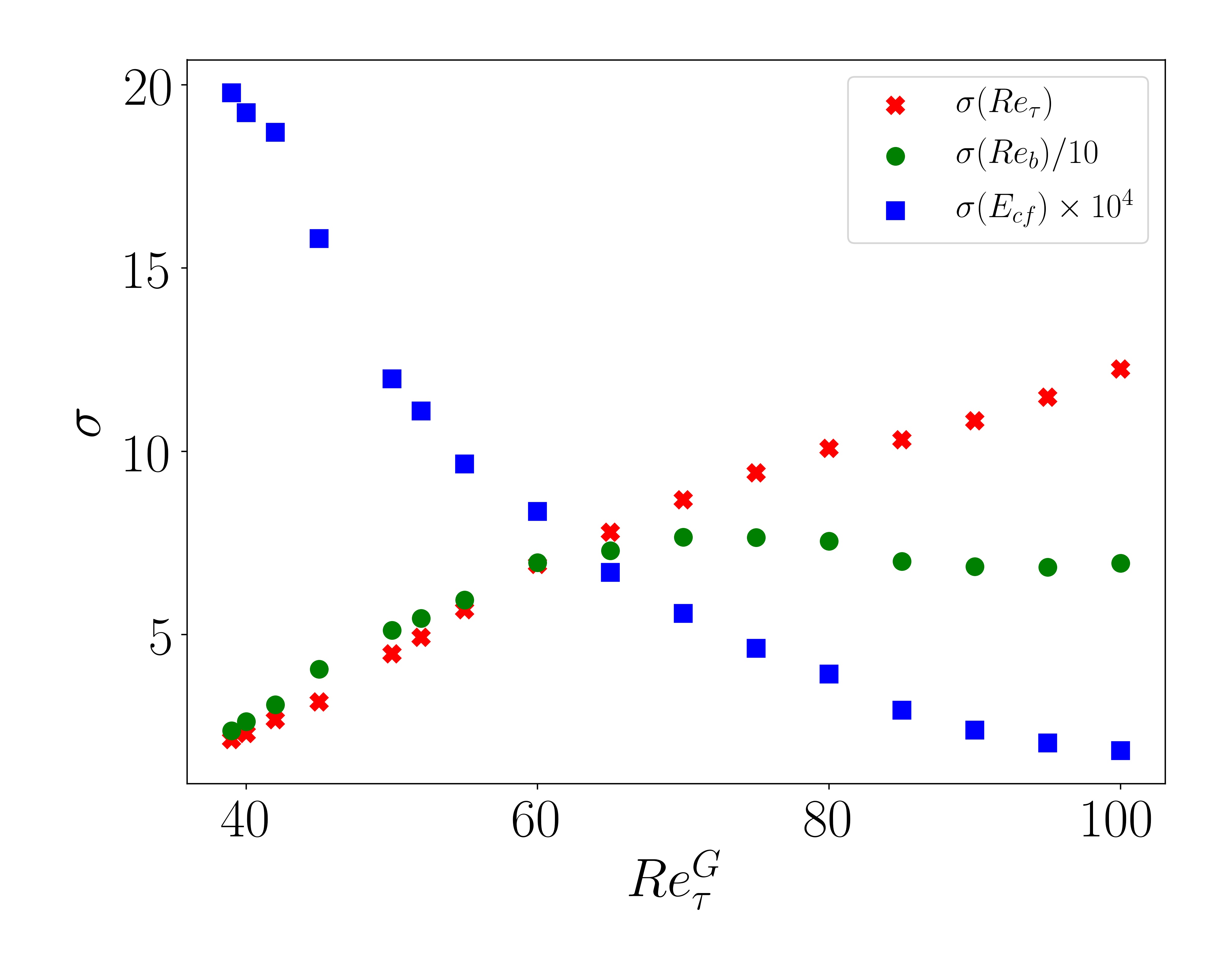}
\caption{}
\label{st2}
\end{subfigure}

\vspace{6pt}
\hfill
\centering
\begin{subfigure}[b]{0.47\textwidth}
\centering
\includegraphics[width=\textwidth]{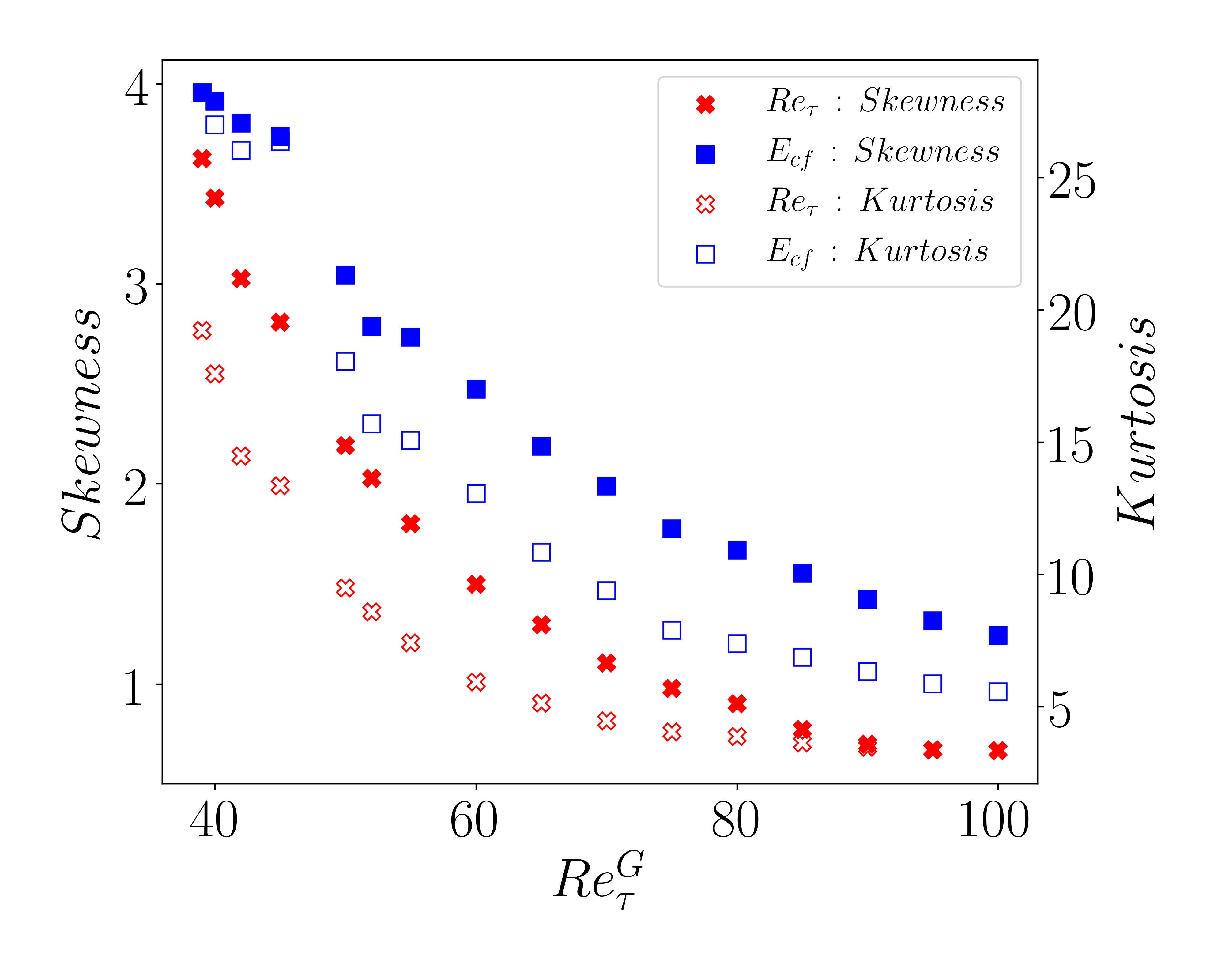}
\caption{}
\label{st3}
\end{subfigure}
\hfill
\begin{subfigure}[b]{0.47\textwidth}
\centering
\includegraphics[width=\textwidth]{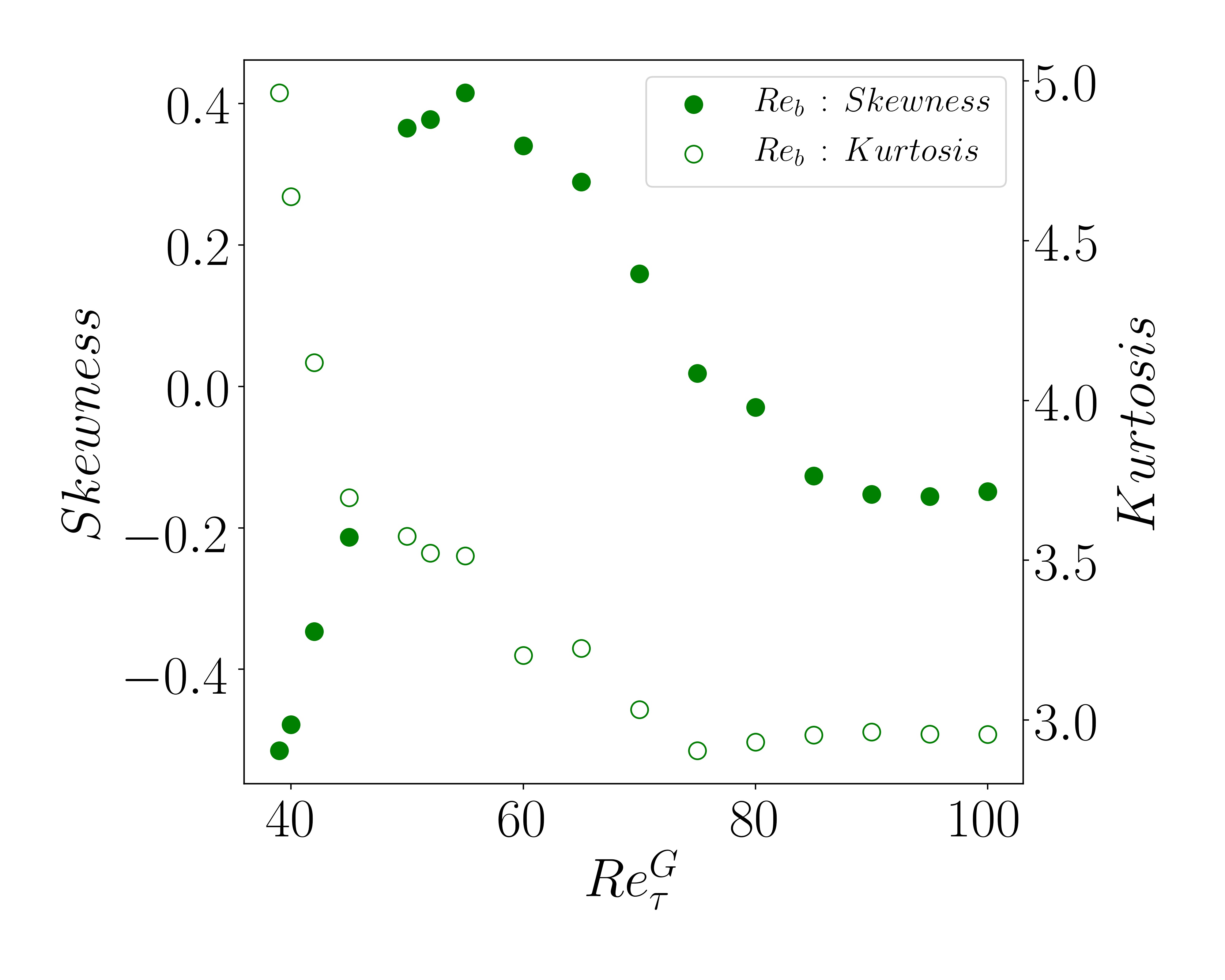}
\caption{}
\label{st4}
\end{subfigure}

\caption{\textit{Cont}.}
\end{figure}

\begin{figure}[H]\ContinuedFloat
\centering
\setcounter{subfigure}{4}

\begin{subfigure}[b]{\textwidth}
\centering
\includegraphics[width=0.5\textwidth]{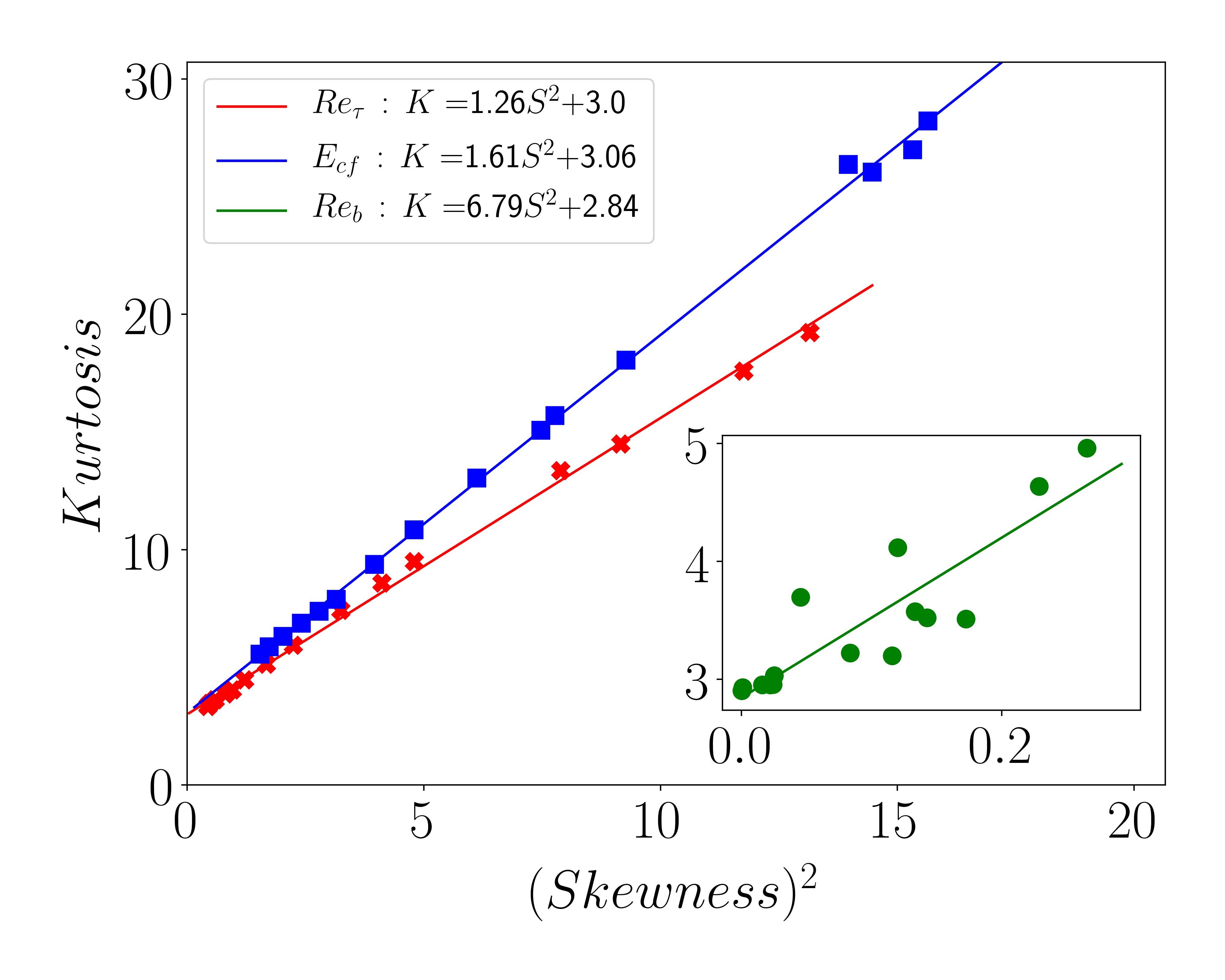}
\caption{}
\label{st5}
\end{subfigure}

\caption{ (\textbf{a}) Mean values ($x_m$) of $Re_b$ and $Re_{\tau}$. (\textbf{b}) Variation of the Standard deviation ($\sigma$) of $Re_\tau$ (red), $Re_b$ (green), $E_{cf}$ (blue) (indicated in the legend) vs. $Re^{\scaleto{G}{4pt}}_{\tau}$. The $\sigma(Re_b)$ and $\sigma(Re_{\tau})$ are scaled as indicated in the legend in order make them comparable. (\textbf{c}) Variation of Skewness ($y$-axis on left, filled symbols) and kurtosis (right $y$-axis, open symbols) vs. $Re^{\scaleto{G}{4pt}}_{\tau}$ for the observables $Re_{\tau}$ (red) and $E_{cf}$ (blue) (\textbf{d}) Variation of Skewness (left $y$-axis on the left, filled symbol) and kurtosis ($y$-axis on right, open symbols) vs. $Re^{\scaleto{G}{4pt}}_{\tau}$ for the observable $Re_b$ (green). (\textbf{e}) Kurtosis vs. squared skewness for $Re_\tau$ (red), $Re_b$ (green, inset), $E_{cf}$ (blue).}
\label{stat}
\end{figure}

\section{Discussion} \label{sdis}

The present simulations of the transitional regime of pPf confirm and extend previously documented knowledge, such as the constancy of $C_f$ in the patterning regime and the variation of the band orientations close to the transition point.

The statistical analysis of the distribution of laminar gaps reveals that the distributions are exponentially tailed over  the entire parameter range $39 \leq Re^{\scaleto{G}{4pt}}_{\tau} \leq 100 $, demonstrating that even the value $Re^{\scaleto{G}{4pt}}_{\tau}=39$ remains away from any sort of critical regime, which would be marked by algebraic distributions. This is consistent with the existing estimation of the location of the transitional critical point $Re_{cl} \approx 660$ \cite{Tao2018extended,paranjape2019thesis}, which translates to $Re^{\scaleto{G}{4pt}}_{\tau} \approx 36$.
The entire patterning regime should thus be seen as bona fide spatio-temporal intermittency,  with the critical behavior and transition point being relegated to values of $Re^{\scaleto{G}{4pt}}_{\tau}<39$. Exploring the statistics of the flow closer to the critical point would require even larger domains and longer observation times. Such an investigation is outside the scope of the current study.

The orientation of the bands in the patterning regime for $60 \leq Re^{\scaleto{G}{4pt}}_{\tau} \leq 90$ ($1800 \leq Re_{cl} \leq 4050$) is essentially constant, with an angle $\theta= 25^{\circ} \pm 2.5^{\circ}$. This validates the choice of $\theta = 24^{\circ}$ as a suitable value in the numerical approach of Tuckerman {et al.} \cite{Barkley2005computational,Tuckerman2014turbulent,gome2020statistical}, where slender computational domains are tilted at a chosen value of the angle. However, this angle of $24^{\circ}$ no longer fits the mean orientation of the independent turbulent bands in the lower range $Re^{\scaleto{G}{4pt}}_{\tau} \leq 60$ ($Re_{cl} \leq 1800$), where the orientation of the bands increases by a factor close to two, with $\theta \approx 40^{\circ}$ for $Re^{\scaleto{G}{4pt}}_{\tau}=39$.

We confirm the observation of a constant $C_f$ in the patterning regime, which also implies \mbox{$\widebar{\left<Re_{\tau}\right>} \sim  \widebar{\left<Re_b\right>}$}, as reflected in Figure \ref{stat}a. This constant value of $C_f$ in the transitional regimes further enforces the long lasting analogy with first order phase transitions \cite{Pomeau1986front}, for which the thermodynamic parameter conjugated to the order parameter remains constant while the system evolves from one homogeneous phase to the other, when a suitable control parameter is varied.  At the mean-field level, a trademark of phase coexistence, is then the presence of a bimodal distribution of the order parameter in the coexistence regime. Capturing this bi-modality is however known as being a challenge, even~in simulations of standard equilibrium systems: first, not all protocols allow for observing the phase coexistence; second, the order parameter must be coarse-grained on appropriate length-scales as compared to the correlation lengths such that non-mean-field effect do not dominate \cite{Binder1984}. More than often, the bi-modality of the order parameter distribution is replaced by a mere concavity and a large kurtosis. If the two phases have very different fluctuations, as is the case here, one also expects a strong skewness of the distribution. Our observations extend the analogy, already reported at the level of the mean observable, to their fluctuations. However, a lot remain to be done to further exploit this analogy, in particular by making more precise what the relevant order and control parameters are. Let~us~stress that whether the analogy with a first order transition is valid or not, it does not preclude the dynamics at the spinodals from obeying a critical scenario, such as directed percolation close to the laminar phase spinodal \cite{Barkley2016theoretical} and a modulated instability of the turbulent flow close to the turbulent~one~\cite{Prigent2002large}.

Finally, the statistical moments showcased here demonstrate a correlation between the skewness and the kurtosis of both $Re_{\tau}$ and $E_{cf}$. Such a correlation, observed in both the transitional regime and higher Reynolds number turbulence but originally developed for the latter only \cite{jovanovic_statistical_1993}, suggests a universal turbulent character, beyond the mere distinction transitional/featureless.

\section{Conclusions} \label{scon}

The transitional regime of pPf has been investigated numerically in large periodic domains. The~transitional regime is composed of two sub-regimes each demarcated by a distinct behavior. The~\emph{patterning regime} is characterized, for $50 \leq Re^{\scaleto{G}{4pt}}_{\tau} \leq 90$, by a constant value of $C_f \approx 0.01$ and by a propagation downstream at approximately the mean bulk velocity $<u_b>$. For lower $Re^{\scaleto{G}{4pt}}_{\tau}$ all the way down to the critical point close to $Re^{\scaleto{G}{4pt}}_{\tau} \leq 36$, independent turbulent bands define a regime analogous to the puff regime of cylindrical pipe flow. The patterns are shown to exhibit a near constant angle of inclination $\theta = 25 ^{\circ} \pm 2.5 ^{\circ}$ for $60 \leq Re^{\scaleto{G}{4pt}}_{\tau} \leq 90$, which increases with reducing $Re^{\scaleto{G}{4pt}}_{\tau}$. Both sub-regimes can be classified as spatiotemporally intermittent, as demonstrated by the exponential tails of the distribution of laminar gaps. The statistics of the local fields $\tau$ and $u_b$ reinforce the feeling that a fruitful analogy with first order phase transitions could be developed, but the later remains to be made more precise and exploited.

\vspace{6pt}

\authorcontributions{Conceptualization, Y.D. and O.D.; methodology, P.V.K., Y.D. and O.D.;  data curation, P.V.K.; original draft preparation, Y.D. and P.V.K.;  visualization, P.V.K.; supervision, Y.D. and O.D. All authors have read and agreed to the published version of the manuscript.}

\funding{This research received no external funding. }

\acknowledgments{This study was made possible using computational resources from IDRIS (Institut du D\'eveloppement et des Ressources en Informatique Scientifique) and the support of its staff. We would like to acknowledge and thank the entire team of \emph{channelflow.ch} for building the code and making it open source. The~authors would also like to thank Takahiro Tsukahara, Kazuki Takeda, Jalel Chergui, Florian Reetz, Rob Poole, Rishav Agrawal, Laurette S. Tuckerman, and Sebastian Gom\'e for valuable discussions and technical input.}

\conflictsofinterest{The authors declare no conflict of interest.}



\reftitle{References}



\end{document}